\documentclass[12pt]{article}

\input epsf.sty

\newcommand{\insertfig}[2]{\mbox{\epsfxsize=#1cm \epsfbox{#2.eps}}}

\newcommand{\Bx}{x_{\rm B}}
\newcommand{\bit}[1]{\mbox{\boldmath$#1$}}
\font\cmss=cmss12 
\def\1{\hbox{{1}\kern-.25em\hbox{l}}}
\def\bfZ{\relax{\hbox{\cmss Z\kern-.4em Z}}}

\begin{document}

\begin{titlepage}

\begin{flushright}
DOE/ER/40762-263 \\ [-2mm] UMD-PP\#03-005
\end{flushright}

\centerline{\large \bf Final state interactions and gauge invariant parton distributions}

\vspace{15mm}

\centerline{\bf A.V. Belitsky, X. Ji, F. Yuan}

\vspace{15mm}

\centerline{\it Department of Physics}
\centerline{\it University of Maryland at College Park}
\centerline{\it MD 20742-4111, College Park, USA}

\vspace{15mm}

\centerline{\bf Abstract}

\vspace{0.5cm}

\noindent

Parton distributions contain factorizable final state interaction effects originating
from the fast-moving struck quark interacting with the target spectators in deeply
inelastic scattering. We show that these interactions give rise to gauge invariance
of the transverse momentum-dependent parton distributions. As compared to previous
analyses, our study demonstrates the existence of extra scaling contributions from
transverse components of the gauge potential at the light-cone infinity. They form a
transverse gauge link which is indispensable for restoration of the gauge invariance
of parton distributions in the light-cone gauge where the gauge potential does not
vanish asymptotically. Our finding helps to explain a number of features observed in
a model calculation of structure functions in the light-cone gauge.

\vspace{5.5cm}



\end{titlepage}

\section{Parton model and QCD}

Hadron structure functions, measurable in deeply inelastic scattering,
are genuine physical observables which provide direct access to the
microscopic constituents of matter and their intricate interaction
dynamics. In the naive parton model \cite{Fey71}, the structure function
is expressed in terms of a probability density $q(x)$ to find a parton
of a specific flavor with a certain fraction $x$ of the parent hadron's
momentum. The underlying probabilistic picture for the scattering process
relies on the fact that the constituents in a hadron boosted to the infinite
momentum frame behave as a collections of noninteracting quanta due to
time dilation. This simple and intuitive description of hard reactions
has found its firm foundation in rigorous field theoretical approach based on
asymptotically free Quantum Chromodynamics (QCD). The result is factorization
theorems which separate incoherent contributions responsible for physics of
large and small distances involved in hard reactions in a universal and
controllable manner: The physical observables such as structure functions are
calculated as a convolution of QCD parton distributions in the hadrons and
parton scattering cross sections. The parton model result arises as a lowest
order term in the expansion in the coupling constant and inverse power of the
hard momentum transfer of QCD factorization formulas.

The QCD quark distribution follows from the factorization theorem in deeply
inelastic scattering \cite{ColSop82}
\begin{equation}
\label{CollinearPDFs}
q (x)
=
\frac{1}{2}
\int \frac{d \xi_-}{2 \pi}
{\rm e}^{- i x \xi_-}
\langle P |
\bar\psi (\xi_-, \bit{0}) \gamma_+ [\xi_-, \bit{0}; 0, \bit{0}] \psi (0, \bit{0})
| P \rangle
\, ,
\end{equation}
where
\begin{eqnarray*}
[ \xi_-, \bit{\xi}; 0, \bit{\xi} ]
\equiv
P \exp \left(
- i g \int^{\xi_-}_0 d \xi_- A_+ ( \xi_-, \bit{\xi})
\right) \, ,
\end{eqnarray*}
is the gauge link between the quark fields, which arises from final state
interactions between the struck quark and the target spectators. This interaction
does not ruin factorization and is in fact much needed to maintain gauge invariance.
On the other hand, the presence of this gauge link seems to spoil the
interpretation of $q(x)$ as a pure quark distribution, as the bilocal operator in
the above expression is not obviously a quark number operator. The probabilistic
interpretation is expected to hold only in the light-cone gauge \cite{BasNarSol91,Lei94},
\begin{equation}
\label{LCgaugeCondition}
A_+ = 0
\, ,
\end{equation}
since only the physical degrees of freedom remain with this choice. In this
special gauge, the light-cone gauge link in the parton distribution in Eq.\
(\ref{CollinearPDFs}) formally vanish. The final state interaction effects
between the outgoing parton and the target spectators have seemingly been removed
and parton distributions depend only on the ground state wave functions. In
particular, $q(x)$ can be calculated as the sum of the proton light-cone
wave functions squared. The above observation is the basis for standard
folklore that QCD parton distributions represent the density of partons in the
target hadron or nucleus.

Recently, however, the authors in Ref.\ \cite{BroHoyMarPeiSan01} questioned
the validity of the standard result that the structure functions are given
by the densities of partons in a target hadron. They argued that final state
interactions between the struck quark and the target spectators have an
intrinsic physical content and cannot be simply disposed off by a clever gauge
choice. In the light-cone gauge, they found that although the final state
interactions between the struck quark and the target spectators cancel after
the transverse-momentum integrations, the effect remains in the interaction
between the target spectators. They concluded on this ground that the parton
distribution in the light-cone gauge is not given by Eq.\ (\ref{CollinearPDFs}):
As such, structure functions cannot be just parton densities in an initial state
hadron, which presumably does not know anything about final state interactions.

The light-cone gauge analysis is complicated by the fact that the condition
$A_+ = 0$ alone does not remove all gauge freedom: $x_-$-independent
gauge transformations are still possible since they do not change the gauge
condition. This leads to an difficulty in perturbation theory as the gauge
propagator develops singularities in the Fourier conjugated momentum variable.
In order to remove them, one has to fix the gauge completely. One way
to do this is to adopt a specific boundary condition on components of the gauge
potential. However, this may yield non-causal interactions, i.e., the final state
interaction effects can be shifted entirely or partially either before or after
the hard scattering. Moreover, the hadron wave functions do depend on the
choice of the boundary conditions. The effect of the migration of final state
interactions was noticed early in the analysis of shadowing
\cite{MueQiu86}, see also \cite{Mue01}.

Let us briefly summarize the results of this paper in light of findings in Ref.\
\cite{BroHoyMarPeiSan01}. To fully exhibit the final state interaction
effects, the Feynman parton distributions are considered as the integration
of the transverse-momentum dependent parton distribution,
\begin{eqnarray*}
q(x) = \int d^2 \bit{k} \ q (x, \bit{k}) \, .
\end{eqnarray*}
As we already emphasized above, the final state interaction effects in
$q(x, \bit{k})$ arise from the interaction between the active quark after it
was hit by the virtual photon and the target remnant. We calculate it in
the deeply inelastic lepton production of a jet with small transverse momentum
and find
\begin{eqnarray*}
q (x, \bit{k}) = \frac{1}{2}
\int \frac{d \xi_-}{2 \pi} \frac{d^2 \bit{\xi}}{(2 \pi)^2}
{\rm e}^{- i x \xi_- + i \bit{\scriptstyle k} \cdot \bit{\scriptstyle \xi}}
\langle P |
\bar\psi (\xi, \bit{\xi}) [\infty, \bit{\infty}; \xi_-, \bit{\xi}]^\dagger_C
\, \gamma_+
[\infty, \bit{\infty}; 0, \bit{0}]_C \, \psi (0, \bit{0})
| P \rangle \, ,
\end{eqnarray*}
with path ordered exponentials stretched in light-like as well as transverse
directions,
\begin{eqnarray*}
[\infty, \bit{\infty}; \xi_-, \bit{\xi}]_C
\equiv
[\infty, \bit{\infty}; \infty, \bit{\xi}]
[\infty, \bit{\xi}; \xi_-, \bit{\xi}] \, .
\end{eqnarray*}
This definition, which did not assume the vanishing of the vector potential
at infinity, is completely gauge-invariant. In the light-cone gauge
(\ref{LCgaugeCondition}), the gauge link along the light-cone disappears,
$[\infty, \bit{\xi}; \xi_-, \bit{\xi}] = 1$, and we are left with the
link in the transverse direction, $[\infty, \bit{\infty}; \infty, \bit{\xi}]$,
which is crucial to maintain the gauge independence under residual gauge
transformations. The effect of final state interactions can thus be
present even in the light-cone gauge. As is clear from the above equation,
the usual transverse momentum-dependent parton distributions with light-cone
gauge links only are incomplete when they are used in the light-cone gauge,
--- they are no longer gauge invariant under different choices of the gauge
boundary conditions. The transverse link in our new definition is responsible
for the final state interactions which generates the Sivers distribution
function: the parton transverse momentum distributions asymmetry in a
transversely polarized nucleon \cite{JiYua02}. This distribution is
responsible fully for the single transverse-spin asymmetry discovered in
\cite{BroHwaSch02a} and explained in \cite{Col02}. The mechanism which led
to nonvanishing asymmetry was thought to be time-reversal odd and therefore
prohibited in QCD. Our analysis supports the conclusion in \cite{Col02} that
time-reversal does not impose any constraint on the Sivers function and
hence the rich phenomenology of transverse momentum-dependent parton
distributions as related to spin physics, see, e.g.,
\cite{Siv90,Col92,AnsBogMur95,AnsMur98,MulTan95,BoeMul97,BarDraRat01}.

With different choices of boundary conditions one can shift residual final
state interactions from the final to the initial state. For the advanced
boundary condition under which the transverse field $\bit{A}$ vanishes at
$\xi_- = \infty$, the transverse gauge link vanishes as well, and the
transverse momentum distributions are computable solely as a sum of the
light-cone wave functions squared! Where are final state interactions now?
They are now contained in the light-cone wave function due to the boundary
condition on the gauge potential. Since the hadron wave function is a
gauge-dependent object, it may acquire certain effects with the change of
the gauge adopted to solve the bound state equation \cite{BroPauPin97}.
Choosing the light-cone gauge with a specific boundary condition one mimics
final state interactions by a phase $\varphi$ in the wave function,
${\mathit \Psi} = |a| {\rm e}^{i \varphi}$, which reproduces the effects of
the gauge links if a different boundary condition is to be chosen.
Therefore, the light-cone wave functions are special in that they encode not
just the ground state properties of a system but also the final state
interaction phases introduced through the choice of gauge conditions. This
holds even though they represent stable bound state particles. This is not
entirely surprising because final state interactions represented by the gauge
links are quite general in nature, independent of details of the jet. In our
view, this is a feature, rather than a defect, of the light-cone wave functions.

Now return back to the ordinary Feynman distributions. The transverse link at
infinity cancels when integration over $\bit{k}$ is performed. Therefore, independent
of the residual gauge fixing, Eq. (1) is the correct formula for the parton
distribution in any gauge. In the light-cone gauge, the light-cone links reduces
to unity, and the parton distribution can be calculated as the sum over the
amplitude squared of the light-cone wave functions. The effects of final state
interactions are partially cancelled through the transverse momentum integration
and partially included already in the light-cone wave functions. So the parton
distributions will be uniquely determined by the light-cone wave functions, $q (x) \sim
|{\mathit \Psi}|^2$, and may be interepreted as the parton density
in the spirit of the conventional Feynman parton model. We
emphasize again here that the conclusion is independent of the residual gauge
fixing in the light-cone gauge.

Therefore the parton distributions do contain factorizable
final state interactions. However, they can be completely hidden in the
hadron wave functions if one chooses light-cone gauge and the advanced
boundary condition. Then the parton distributions become densities
in the sense that they can be calculated as the sum over the light-cone
wave functions squared. As an example that the light-cone wave functions
contain final-state intereaction effects, the light-cone amplitudes
are complex when the advanced boundary condition is chosen \cite{JiMaYu02},
and they do not possess simple time-reversal symmetry properties.
Hence some single spin asymmetries can be attributed to, albeit
in a non-causal way, the properties
of the bound-state wave functions!

A complementary view of the gauge links
is to regard the (outgoing/incoming) ``physical parton" as a free parton
field with an attached gauge link extending from the parton
position to (positive/negative) infinity. This ``physical parton" field
is gauge invariant and contains all the initial/final state
interaction effects. The parton distributions are then the densities
of these ``physical partons" in a bound state.

The outline of our presentation is as follows. In the next
section we introduce objects of study and spell out our conventions. Next we
address the issue of the gauge invariance of transverse momentum-dependent
parton distribution in deeply inelastic scattering (DIS) by deriving the gauge
link. As we stated above, it extends along the light-cone direction and
is accompanied by an extra scaling contribution from the transverse components
of the gauge field at light-cone infinity. The latter contribution would not
be present for local gauge potentials, however, it affects the distribution
in the light-cone gauge which generates non-local and non-causal interactions.
In section \ref{PDFinQEDmodel}, we calculate parton distributions in a simple
scalar QED model of Ref.\ \cite{BroHoyMarPeiSan01} and demonstrate the effect
of the final state interactions in the interplay with boundary conditions
imposed on the gauge potential. We demonstrate the utter equivalence of different
prescription for the light-cone pole on the amplitude level provided the
transverse gauge link is accounted for. We show that with the advanced boundary
conditions the final state interactions vanish and the parton distribution are
defined solely by hadronic wave functions. In section \ref{Dipoles}, we show
the equivalence of the dipole picture of deeply inelastic scattering in the
aligned-jet kinematics, as discussed in Ref.\ \cite{BroHoyMarPeiSan01},
to the twist-two structure functions. In section \ref{DYversusDIS}, we
address transverse momentum-dependent parton distribution functions in
the Drell-Yan (DY) process and discuss violation of the naive universality
with DIS distribution functions. A procedure for the
complete fixing of the residual gauge freedom in the light-cone gauge
(\ref{LCgaugeCondition}) by imposing a set of boundary conditions is
given in the appendix \ref{Residual}.

\section{Definitions}

The main object of our study will be the structure functions which
parametrize the hadronic part of the deeply inelastic scattering cross
section and expressed, in turn, by the absorptive part of the forward
Compton scattering amplitude
\begin{equation}
W_{\mu\nu} = \frac{1}{2 \pi} \Im{\rm m} T_{\mu\nu} \, .
\end{equation}
The latter is a matrix element of the chronological product of quark
electromagnetic currents $j_\mu = \sum_q Q_q \bar \psi_q \gamma_\mu \psi_q$,
\begin{equation}
\label{ForwardCompton}
T_{\mu\nu} = i \int d^4 \xi \, {\rm e}^{i q \cdot \xi}
\langle P | T \left\{ j_\mu (\xi) j_\nu (0) \right\} | P \rangle \, .
\end{equation}
Using the completeness of hadronic states $\sum_N | N \rangle \langle N | = 1$,
with a state $| N \rangle$ consisting of $N$ particles with the total momentum
$P_N = \sum_{k = 1}^N p_k$ and certain quantum numbers, one gets
\begin{eqnarray}
W_{\mu\nu}
\!\!\!&=&\!\!\! \frac{1}{4 \pi}
\sum_N \int \prod_{k = 1}^N \frac{d^4 p_k}{(2 \pi)^4}
(2 \pi) \delta_+ \left( p_k^2 - m_k^2 \right)
(2 \pi)^4 \delta^{(4)} \left( \sum_{l = 1}^N p_l - P - q \right)
\nonumber\\
&&\qquad\qquad\qquad\qquad\qquad\qquad\qquad\quad\times
\langle P | j_\mu (0) | p_1, \cdots, p_N \rangle
\langle p_1, \cdots, p_N | j_\nu (0) | P \rangle \, ,
\end{eqnarray}
where $\delta_+ (p^2 - m^2) = \theta (p_0) \delta (p^2 - m^2)$ imposes
the on-mass-shell condition for an outgoing particle and the summation over
$N$ involves the summation over the number of particles populated the final
states as well as their quantum numbers.

For an unpolarized target, the hadronic tensor admits the following
decomposition into the structure functions
\begin{equation}
W_{\mu\nu}
=
\left( - g_{\mu\nu} + \frac{q_\mu q_\nu}{q^2} \right)
F_1 (\Bx, Q^2)
+
\frac{1}{P \cdot q}
\left( P_\mu - q_\mu \frac{P \cdot q}{q^2} \right)
\left( P_\nu - q_\nu \frac{P \cdot q}{q^2} \right)
F_2 (\Bx, Q^2) \, ,
\end{equation}
which depend on the photon virtuality $Q^2 = - q^2$ and the Bjorken variable
$\Bx = Q^2/(2 q \cdot P)$. In the Bjorken limit $Q^2 \to \infty$ with $\Bx =$
fixed, these functions are expressed,
\begin{equation}
F_1 (\Bx, Q^2) = \frac{1}{2 \Bx} F_2 (\Bx, Q^2)
=
\frac{1}{2} \sum_q Q_q^2 \left( f_q (\Bx, Q^2) +  f_{\bar q} (\Bx, Q^2) \right)
\, ,
\end{equation}
in terms of QCD quark $f_q$ and antiquark $f_{\bar q}$ distribution functions,
at leading order of QCD perturbation theory and neglecting power-suppressed
contributions. The dependence of $f_{q, \bar q}$ on $Q^2$ arises only via the
renormalization of the underlying composite operators and is logarithmic. In
our subsequent considerations this will be irrelevant and, in order to simplify
formulas, we neglect it in parton distributions. We also drop the summation
over the quark flavors and set the quark charge to unity, $Q_q = 1$.

The Bjorken kinematics sets the microscopic dynamics on the light-cone, since
the photon four-vector has large $q_- = Q^2/(2 \Bx)$ and small $q_+
= - \Bx$ component so that the integrand in Eq.\ (\ref{ForwardCompton}) does
not oscillate provided $\xi_+ \sim 1/q_-$ is small and $\xi_- \sim 1/q_+$ is
large. In a frame where $q$ does not have the transverse components, the
light-cone decomposition of the external momenta reads
\begin{equation}
q_\mu = - \Bx n^\ast_\mu + \frac{Q^2}{2 \Bx} n_\mu
\, \qquad
P_\mu = n^\ast_\mu + \frac{M^2}{2} n_\mu \, .
\end{equation}
We introduced the light-like vectors $n_\mu$ and $n^\ast_\mu$ such that
\begin{equation}
n^2 = n^{\ast 2} = 0 \, , \qquad n \cdot n^\ast = 1 \, .
\end{equation}
For an arbitrary four-vectors $v_\mu$, we define
\begin{equation}
\label{Sudakov}
v_+ \equiv v \cdot n \, , \quad
v_- \equiv v \cdot n^\ast \, , \quad
v \cdot u = v_+ u_- + v_- u_+ + v_\perp \cdot u_\perp \, .
\end{equation}
We use Euclidean notations for transverse two-dimensional space with the
metric $\delta_{\alpha\beta} = - g^\perp_{\alpha\beta} = - ( g_{\alpha\beta}
- n_\alpha n^\ast_\beta - n_\alpha^\ast n_\beta ) = {\rm diag} (1, 1)$, so that
\begin{equation}
v_\perp \cdot u_\perp = - \mbox{\boldmath$v$} \cdot \mbox{\boldmath$u$}
\, .
\end{equation}
Using these definitions one can easily project the hadronic structure functions
via
\begin{equation}
\label{StructureProjection}
F_1 = - \frac{1}{2} g^\perp_{\mu\nu} W_{\mu\nu}
\, , \qquad
F_L = \frac{(2 \Bx)^3}{Q^2} W_{--}
\, ,
\end{equation}
where we defined the function $F_L \equiv F_2 - 2 \Bx F_1$. These structure
functions are related to the absorption cross sections of transversely and
longitudinally polarized photon by the hadron, $F_1 \sim \sigma^{\gamma^\ast}_{T}$
and $F_L \sim \sigma^{\gamma^\ast}_{L}$, respectively. In the Bjorken limit,
$F_1 \gg F_L$ for spin-$1/2$ constituents, while for scalar partons the
inequality is reversed, $F_1 \ll F_L$.

\section{Parton distributions}
\label{DISpdfs}

In this section we demonstrate how a gauge invariant parton densities
arise in QCD description of the structure functions at leading twist.
In particularly, we consider transverse momentum-dependent distributions.
In QCD, the parton distributions are defined as hadronic matrix elements
of quark bilocal operators. Since the parton fields enter at distinct
space-time points, a gauge link is needed to make the operators gauge
invariant. This gauge link is generated in a hard scattering process by
the final state interactions between the struck parton and the target
remnants. Since the struck parton moves with a high energy, its interaction
can be approximated through an eikonal phase. We show that the conventional
light-cone link is not the only scaling contribution.

A classic analysis \cite{ColSop82} demonstrated that non-gauge invariant
quark bilocal correlators representing parton distributions in the
zero-order approximation acquire path-ordered gauge links stretched along
a non-light-like direction, $n_{\rm NL}$, from the position of quarks to
infinity, and restore in this way their gauge invariance. A naive lifting
of the direction to a light-like $n_{\rm NL} = n$ was shown to lead to
severe divergences \cite{ColSop81}, which cancel only for integrated
densities. Although, not implied by Ref.\ \cite{ColSop82}, the following
definition of the transverse-momentum dependent quark distributions has
been unequivocally accepted in the literature
\begin{equation}
\label{PDFliterature}
q (x, \bit{k})
= \frac{1}{2}
\int \frac{d \xi_-}{2 \pi} \frac{d^2 \bit{\xi}}{(2\pi)^2}
{\rm e}^{- i x \xi_- + i \bit{\scriptstyle k} \cdot \bit{\scriptstyle \xi}}
\langle P |
\bar \psi ( \xi_-, \bit{\xi} )
[\infty, \bit{\xi}; \xi_- , \bit{\xi}]^\dagger
\gamma_+
[\infty, \bit{0}; 0, \bit{0}]
\psi(0, \bit{0})
| P \rangle \ ,
\end{equation}
where the path-ordered gauge link extends along the light-cone
\begin{equation}
[ \infty, \bit{\xi}; \xi_-, \bit{\xi} ]
\equiv
P \exp \left(
- i g \int^\infty_{\xi_-} d \xi_- A_+ ( \xi_-, \bit{\xi})
\right) \, .
\end{equation}
Since the coordinate $\xi_+ = 0$, we do not display it, for brevity, here
and the following presentation. As we will see, Eq.\ (\ref{PDFliterature})
is true only in a class of gauges where the gluon potential vanishes at
$\xi_- = \infty$. Note that the integrated parton distribution is just
\begin{equation}
\label{LightConeDistribution}
q(x) = \int d^2 \bit{k} \, q (x, \bit{k}) \, ,
\end{equation}
and generates quark and antiquark distributions $q (x) = f_q (x) \theta (x)
- f_{\bar q} \theta (- x)$.

\begin{figure}[t]
\begin{center}
\mbox{
\begin{picture}(0,105)(200,0)
\put(0,1){\insertfig{13}{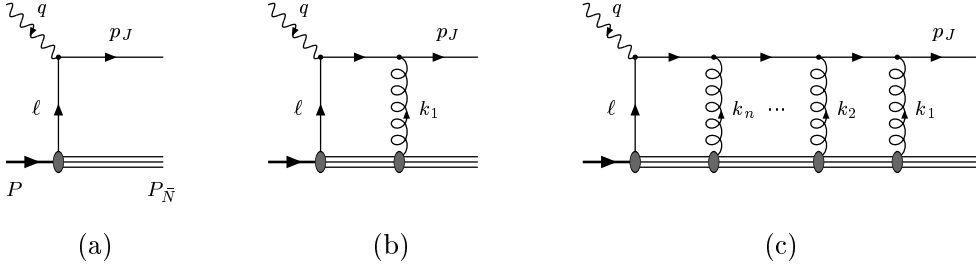}}
\end{picture}
}
\end{center}
\caption{\label{LinkDIS} Multi-gluon attachments to the struck quark in
deeply inelastic scattering which factorize to form the path-ordered
exponential.}
\end{figure}

For the photon scattering on a single parton with momentum $\ell$, we have
for the hadronic tensor
\begin{equation}
\label{DIStensorOne}
W_{\mu\nu} = \frac{1}{4 \pi} \sum_{\bar N} \int \frac{d^4 p_J}{(2 \pi)^4}
(2 \pi) \delta_+ (p_J^2)
(2 \pi)^4 \delta^{(4)} \left( P_{\bar N} + p_J - P - q \right)
\langle P | j_\mu (0) | p_J, \bar N \rangle
\langle p_J, \bar N | j_\nu (0) | P \rangle \, ,
\end{equation}
where $p_J$ is the momentum of the observed final state quark (jet) after
multiple rescattering with spectators $\bar N$ in the target fragments. The
tree scattering amplitude corresponding to Fig.\ \ref{LinkDIS} (a) reads
\begin{equation}
\label{TreeAmplitude}
\langle p_J, \bar N | j_\nu (0) | P \rangle_{(0)}
=
\bar u (p_J) \gamma_\nu \langle \bar N | \psi(0) |P \rangle \, ,
\end{equation}
where $\bar u$ is the Dirac spinor of the scattered quark. Substituting Eq.\
(\ref{TreeAmplitude}) into (\ref{DIStensorOne}) we get, after projection
with (\ref{StructureProjection}), the structure function $F_1 (\Bx, Q^2)$
in the parton model
\begin{equation}
F_1 (\Bx, Q^2)
=
\frac{1}{2} \int d^2 \bit{p}_J \, q (\Bx, \bit{p}_J) \, ,
\end{equation}
in terms of the parton distribution
\begin{equation}
q (x, \bit{k})
= \frac{1}{2}
\int \frac{d \xi_-}{2 \pi} \frac{d^2 \bit{\xi}}{(2\pi)^2}
{\rm e}^{- i x \xi_- + i \bit{\scriptstyle k} \cdot \bit{\scriptstyle \xi}}
\langle P |
\bar \psi ( \xi_-, \bit{\xi} )
\gamma_+
\psi(0, \bit{}0)
| P \rangle \, .
\end{equation}
As it stands the above correlation is not gauge invariant.

\subsection{Light-cone gauge link}

Consider now the contribution from the diagram in Fig.\ \ref{LinkDIS}
(b). The one-gluon amplitude reads
\begin{equation}
\label{OneGluonAmplitude}
\langle p_J, \bar N | j_\nu (0) | P \rangle_{(1)}
=
g \bar u (p_J) \int \frac{d^4 k_1}{(2 \pi)^4}
\langle \bar N |
{\not\!\! A} (k_1) S_0 (p_J - k_1) \gamma_\nu \psi(0)
| P \rangle \, ,
\end{equation}
where the free quark propagator is
\begin{equation}
\label{FreeFermionPropMomentum}
S_0 (k) = \frac{\not\! k}{k^2 + i 0} \, .
\end{equation}
Momentum conservation gives
\begin{eqnarray*}
p_J = q + \ell + k_1 \, ,
\end{eqnarray*}
so that the on-mass-shell condition $p_J^2 = 0$ results into
\begin{eqnarray*}
\ell_+ + k_{1 + } = \Bx \, ,
\end{eqnarray*}
in the Bjorken limit $q_- \to \infty$. By expanding all vectors in Sudakov
components (\ref{Sudakov}), one can easily keep track of leading contributions
in the scaling limit. For instance, the struck quark propagator contains
a scaling term,
\begin{eqnarray}
S_0 (p_J - k_1) \approx
\frac{q_- \gamma_+}{2 q_- (\ell_+ - \Bx) + i 0}
= - \frac{1}{2} \frac{\gamma_+}{k_{1 + } - i 0} \, ,
\end{eqnarray}
with the remainder naively suppressed by extra powers of $1/q_-$.

Making the light-cone decomposition of the gluon field
\begin{equation}
A_\mu = n^\ast_\mu A_+ + n_\mu A_- + A^\perp_\mu \, ,
\end{equation}
one notices that the leading twist contribution comes from the first term
on the right-hand side only, since the second one vanishes due to nilpotence
of $\gamma_-^2 = 0$, while the transverse component of the gauge field is of
twist-three. Since the leading component in the quark density matrix
$u \otimes \bar u$ comes from the large term $q_- \gamma_+$ we are allowed
to replace $\bar u \gamma_- \gamma_+$ by $2 \bar u$ on the right-hand side
of Eq.\ (\ref{OneGluonAmplitude}). Finally, integrating with respect to $k_1$
we get
\begin{equation}
\label{FirstTermPath}
\int \frac{d^4 k_1}{(2 \pi)^4} \frac{1}{k_{1+} - i 0} A (k_1)
= i \int_{- \infty}^\infty d \xi_- \theta (\xi_-) A_+ (\xi_-, 0, \bit{0})
\, .
\end{equation}
Using these results, we find that the amplitude in one-gluon approximation
differs from the three-level result (\ref{TreeAmplitude}) only by an extra
factor, namely,
\begin{equation}
\label{LightLinkFirst}
\langle p_J, \bar N | j_\nu (0) | P \rangle_{(1)}
= \bar u (p_J) \gamma_\nu
\langle \bar N |
(- i g) \int_0^\infty d \xi_- A_+ (\xi_-, 0, \bit{0})
\psi (0)
| P \rangle \, .
\end{equation}

These considerations are generalizable in a straightforward manner to an
arbitrary order. Namely, for $n$-gluon exchange, see Fig.\ \ref{LinkDIS} (c),
we have
\begin{eqnarray}
\langle p_J, \bar N | j_\nu (0) | P \rangle_{(n)}
\!\!\!&=&\!\!\!
g^n \bar u (p_J) \int \prod_{i = 1}^n \frac{d^4 k_i}{(2 \pi)^4}
\langle \bar N |
{\not\!\! A} (k_1) S_0 (p_J - k_1)
{\not\!\! A} (k_2) S_0 (p_J - k_1 - k_2)
\cdots \nonumber\\
&&\qquad\qquad\qquad\qquad\ \ \times
{\not\!\! A} (k_n) S_0 \left( p_J - \sum_{j = 1}^n k_j \right)
\gamma_\nu \psi(0)
| P \rangle \, .
\end{eqnarray}
Making repeatedly the same set of approximations as above, one simplifies
the Dirac structure to
\begin{eqnarray}
\langle p_J, \bar N | j_\nu (0) | P \rangle_{(n)}
\!\!\!&=&\!\!\! (- g)^n \int \prod_{i = 1}^n \frac{d^4 k_i}{(2 \pi)^4}
\frac{1}{k_{1+} - i 0} \frac{1}{k_{1+} + k_{2+} - i 0}
\dots
\frac{1}{\sum_{j = 1}^n k_{j + } - i 0}
\nonumber\\
&&\qquad\qquad\qquad\ \times
\bar u (p_J) \gamma_\nu
\langle \bar N |
A_+ (k_1) A_+ (k_2) \cdots A_+ (k_n) \psi (0)
| P \rangle \, ,
\end{eqnarray}
where we have used the momentum conservation $\ell_+ + \sum_{i = 1}^n k_{i + }
= \Bx$. The calculation of the momentum integrals is trivial and gives, similarly
to Eq.\ (\ref{FirstTermPath}),
\begin{eqnarray}
&&\int \prod_{i = 1}^n \frac{d^4 k_i}{(2 \pi)^4}
\frac{1}{k_{1+} - i 0} \frac{1}{k_{1+} + k_{2+} - i 0}
\dots
\frac{1}{\sum_{j = 1}^n k_{j + } - i 0}
A_+ (k_1) A_+ (k_2) \cdots A_+ (k_n)
\nonumber\\
&&\qquad\qquad\qquad
= i^n \int_{- \infty}^\infty
\prod_{i = 1}^n d \xi_{i+} \theta (\xi_{i-} - \xi_{(i + 1)-})
A_+ (\xi_{1 -}) A_+ (\xi_{2 -}) \cdots A_+ (\xi_{n -})
\, ,
\end{eqnarray}
where $\xi_{(n + 1)-} = 0$. Thus,
\begin{eqnarray}
\langle p_J, \bar N | j_\nu (0) | P \rangle_{(n)}
\!\!\!&=&\!\!\! \bar u (p_J) \gamma_\nu
\langle \bar N |
(- i g)^n
\int_0^\infty d \xi_{1-} \int_0^{\xi_{1-}} d \xi_{2-}
\cdots
\nonumber\\
&&\qquad\qquad
\times\int_0^{\xi_{(n - 1)-}} d \xi_{n-}
A_+ (\xi_{1 -}) A_+ (\xi_{2 -}) \cdots A_+ (\xi_{n -})
\psi (0)
| P \rangle \, .
\end{eqnarray}
Here, one immediately recognizes the $n$th term in the expansion of
the path-ordered exponential. Therefore, for the amplitude resummed
to all orders, one gets
\begin{equation}
\label{LightLikeLinkAmplitude}
\langle p_J, \bar N | j_\nu (0) | P \rangle
=
\sum_{n = 0}^\infty
\langle p_J, \bar N | j_\nu (0) | P \rangle_{(n)}
=
\bar u (p_J) \gamma_\nu
\langle \bar N |
P \exp \left( - i g \int_0^\infty d \xi_- A_+ (\xi_-) \right) \psi (0)
| P \rangle \, .
\end{equation}
Substitution of this result into the hadronic tensor yields, indeed, the
conventional quark distribution (\ref{PDFliterature}).

\subsection{Transverse gauge link}
\label{TransverseGaugeLink}

In the previous section we were not quite accurate and actually
omitted contributions which scale in the Bjorken region. These
terms survive only at a point of the momentum space and are, normally,
assumed to be vanishing. However, this is not the case for all
gauge potentials. And a counter-example is provided by the light-cone
gauge where the potential is singular at the very same point.

In the present circumstances it is more convenient to work in a frame
similar to the Drell-Yan frame as was done previously in Refs.\
\cite{MenOlnSop92,Col92}. Namely, the four-momentum of the current jet
$p_J$ is light-like, $p_J^2 = 0$, and can be chosen as one of the
light-like vectors with the other one still fixed by the hadron momentum
$P_\mu = n^\ast_\mu$. Thus, we define
\begin{equation}
\tilde n_\mu \equiv \frac{p_{J\mu}}{p_J \cdot n^\ast} \, .
\end{equation}
Obviously, $\tilde n^2 = n^{\ast 2} = 0$ and $n \cdot \tilde n = 1$.
We decompose all Lorentz tensors via light-cone and transverse components
defined in this basis
\begin{equation}
v_\mu = \tilde n_\mu v_- + n^\ast_\mu v_+ + v^\perp_\mu \, ,
\end{equation}
and keep the same $+$ index for contractions with $\tilde n$, $v_+ \equiv
\tilde n \cdot v$. In the Bjorken limit $q_- \to \infty$, the difference
between $p_{J-}$ and $q_-$ is negligible and, therefore, both frames, used
before and here, coincide.

Let us discuss first one-gluon exchange contribution in Eq.\
(\ref{OneGluonAmplitude}). By looking at the denominator of the quark
propagator one immediately notices that the scaling contribution in
the Bjorken limit $p_{J -} \to \infty$,
\begin{equation}
\frac{1}{(p_J - k_1)^2 + i 0}
\approx
- \frac{1}{2 p_{J -} k_{1 +} + \bit{k}_1^2 - i0} \, ,
\end{equation}
arises not only when one extracts the large $p_{J-}$ component from the
numerator but also when $k_{1+} \sim 1/p_{J -} \to 0$ and keeps finite
contributions in the numerator. A simple algebra gives
\begin{eqnarray}
\bar u (p_J) \gamma_\mu ({\not\! p}_J - {\not\! k}_1) \gamma_\nu
\langle \bar N | A_\mu (k_1) \psi (0) | P \rangle
\!\!\!&\approx&\!\!\!
2 p_{J -} \bar u (p_J) \gamma_\nu
\langle \bar N | A_+ (k_1) \psi (0) | P \rangle
\nonumber\\
&-&\!\!\!
\bar u (p_J) \bit{\gamma}_\alpha {\not\! \bit{k}}_1 \gamma_\nu
\langle \bar N | \bit{A}_\alpha (k_1) \psi (0) | P \rangle
\, .
\end{eqnarray}
where we have used the fact that, when keeping the transverse part of the
gluon field, ${\not\! p}_J$ can now be pushed through ${\not\!\! \bit{A}}$,
giving zero when acting on the on-shell spinor due to the equation of motion
$\bar u (p_J) {\not\! p}_J = 0$. Thus, we find
\begin{eqnarray}
\label{Numerator}
\bar u (p_J)
\langle \bar N |
{\not\!\! A} (k_1) S_0 (p_J - k_1) \gamma_\nu \psi (0)
| P \rangle
\!\!\!&\approx&\!\!\! - \frac{1}{k_{1+} - i 0}
\bar u (p_J) \gamma_\nu
\langle \bar N | A_+ (k_1) \psi (0) | P \rangle
\nonumber\\
&+&\!\!\! \bar u (p_J)
\frac{
\bit{\gamma}_\alpha {\not\! \bit{k}}_1 \gamma_\nu
}{
2 p_{J -} k_{1 +} + \bit{k}_1^2 - i 0
}
\langle \bar N | \bit{A}_\alpha (k_1) \psi (0) | P \rangle
\, .
\end{eqnarray}
The first term on the right-hand side is a contribution to the conventional
light-cone link. When summed with contribution of multi-gluon exchanges
it results as before into Eq.\ (\ref{LightLikeLinkAmplitude}). We will
drop, therefore, these terms completely and concentrate on effects of
the second kind from transverse components of the gluon field.

As we already emphasized before, in the scaling limit of $p_{J- } \to
\infty$, a finite contribution comes from $k_{1+} = 0$, i.e., when the
exchanged gluon carries no longitudinal momentum. To see this, we
exponentiate the denominator via the Chisholm representation
\begin{equation}
\frac{1}{2 p_{J -} k_{1 +} + \bit{k}_1^2 - i 0}
=
i \int^\infty_0 d \lambda \,
{\rm e}^{- i \lambda
( 2 p_{J -} k_{1+} + \bit{\scriptstyle k}_1^2 - i 0 )}
\, .
\end{equation}
Substituting it into (\ref{Numerator}) and then into Eq.\ (\ref{OneGluonAmplitude}),
we can perform the integrations with respect to $k_{1+}$ and $k_{1-}$ which
merely yield the Fourier transform of the gauge potential in these momentum
components $A_\mu (\xi_- = 1/(2 \lambda p_{J -}), \xi_+ = 0, \bit{k}_1)$.
Thus, in the scaling limit $p_{J -} \to \infty$, the argument of $A(\xi)$
is set to $\xi_- = \infty$. The integration over $\lambda$ can now be trivially
performed giving the propagator in the transverse space,
\begin{equation}
\label{OneGLuonIntermediate}
\langle p_J, \bar N | j_\nu (0) | P \rangle_{(1)}
=
g \bar u (p_J) \int \frac{d^2 \bit{k}_1}{(2 \pi)^2}
\bit{\gamma}_\alpha \frac{{\not\! \bit{k}}_1}{\bit{k}_1^2 - i 0} \gamma_\nu
\langle \bar N |
\bit{A}_\alpha (\xi_- = \infty, \xi_+ = 0, \bit{k}_1) \psi(0)
| P \rangle \, .
\end{equation}

To proceed further, we note that $\bit{A}_\alpha (\xi = \infty, \xi_+ = 0, \bit{\xi})$
must be a pure gauge\footnote{Generally, a pure gauge field is $A_\mu (x) = g^\dagger
(x) \partial_\mu g (x)$ with a group-valued function $g (x)$. We assume that $g (x)$
represents small gauge transformations which are contractable to the unit element
$g \approx 1 + \phi$. In the Eq.\ (\protect\ref{PureGauge}) we kept only the leading
term in the perturbative expansion.}
\begin{equation}
\label{PureGauge}
\bit{A}_\alpha (\xi_- = \infty, \xi_+ = 0, \bit{\xi})
=
\bit{\nabla}_\alpha \phi (\bit{\xi}) \, ,
\end{equation}
since the field strength vanishes. The Fourier transform of this potential
to the mixed, light-cone coordinate-transverse momentum representation, is
$\bit{A}_\alpha (\xi_- = \infty, \xi_+ = 0, \bit{k}) = i \bit{k}_\alpha
\widetilde \phi (\bit{k})$, with $\widetilde \phi$ being the Fourier transform
of $\phi (\bit{\xi})$. Substituting it into Eq.\ (\ref{OneGLuonIntermediate})
we cancel the denominator $\bit{k}_1^2$, making use of $\bit{\gamma}_\alpha
\bit{\gamma}_\beta + \bit{\gamma}_\beta \bit{\gamma}_\alpha = - 2
\delta_{\alpha\beta}$, and get
\begin{equation}
\label{TransvOneGluon}
\langle p_J, \bar N | j_\nu (0) | P \rangle_{(1)}
= - i g \bar u (p_J) \gamma_\nu
\langle p_J, \bar N | \phi (\bit{0}) \psi (0) | P \rangle \, .
\end{equation}
It is easy to see that $\phi (\bit{0})$ can be represented as a line integral
\begin{equation}
\phi (\bit{0}) = - \int_0^\infty d \bit{\xi} \cdot \bit{A} (\infty, \bit{\xi}) \, .
\end{equation}
The above expression is the first term in the expansion of an additional eikonal
phase to the conventional one (\ref{LightLinkFirst}), which has been neglected
in the literature. Therefore, even in the light-cone gauge $A_+ = 0$, Fig.\
\ref{LinkDIS} (b) does generate a non-zero contribution to the parton
distribution.

The momentum space procedure just outlined cannot be easily extended beyond
single-gluon exchange. Therefore, the strategy will be to transform all factors
in the integrand of Eq.\ (\ref{OneGLuonIntermediate}) into the transverse
coordinate space and do all manipulations there. To this end, we use
\begin{equation}
\frac{\bit{k}_\alpha}{\bit{k}^2}
= - \frac{i}{2 \pi} \int d^2 \bit{\xi} \,
{\rm e}^{i \bit{\scriptstyle k} \cdot \bit{\scriptstyle \xi}}
\frac{\bit{\xi}_\alpha}{\bit{\xi}^2}
= - \frac{i}{2 \pi} \int d^2 \bit{\xi} \,
{\rm e}^{i \bit{\scriptstyle k} \cdot \bit{\scriptstyle \xi}}
\, \bit{\nabla}_\alpha \ln |\bit{\xi}|
\, ,
\end{equation}
where we do not display a mass parameter which makes the argument of the
logarithm dimensionless. Thus, the right-hand side of Eq.\
(\ref{OneGLuonIntermediate}) can be equivalently written as
\begin{equation}
- i g \bar u (p_J)
\bit{\gamma}_\alpha \bit{\gamma}_\beta \gamma_\nu
\int \frac{d^2 \bit{\xi}}{2 \pi} \,
\bit{\nabla}_\beta \ln |\bit{\xi}| \,
\langle \bar N |
\bit{\nabla}_\alpha \phi (\bit{\xi}) \, \psi(0)
| P \rangle \, .
\end{equation}
Next, one integrates by parts so that both derivatives act on the logarithm,
$\bit{\nabla}_\alpha \bit{\nabla}_\beta \ln |\bit{\xi}|$. This is a symmetric
tensor in its two-dimensional Lorentz indices so that the contracted Dirac
matrices can be also symmetrized and reduced to the Kronecker symbol
$\delta_{\alpha\beta}$ by means of the Cliford algebra relation. Finally, since
$\ln |\bit{\xi}|$ is a two-dimensional Green function, one has
$\bit{\nabla}^2 \ln |\bit{\xi}| = 2 \pi \delta^{(2)} (\bit{\xi})$. These
manipulations lead to the result in Eq.\ (\ref{TransvOneGluon}).

\begin{figure}[t]
\begin{center}
\mbox{
\begin{picture}(0,95)(80,0)
\put(0,1){\insertfig{6}{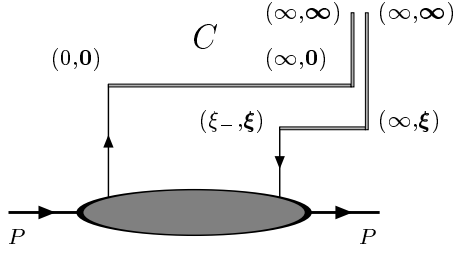}}
\end{picture}
}
\end{center}
\caption{\label{LinksPDFtransverse} Graphical representation of transverse
momentum-dependent parton distribution in the coordinate space with gauge
links endowing nonlocal product of quark fields with gauge invariance.}
\end{figure}

For $n$-gluon exchanges, we have the following contribution from the transverse
components of the gluon field, similarly to Eq.\ (\ref{OneGLuonIntermediate}),
\begin{eqnarray}
\langle p_J, \bar N | j_\nu (0) | P \rangle_{(n)}
\!\!\!&=&\!\!\! g^n \bar u (p_J)
\langle \bar N |
\int \frac{d^2 \bit{k}_1}{(2 \pi)^2}
{\not\!\! \bit{A}} (\infty, \bit{k}_1)
\frac{{\not\! \bit{k}}_1}{\bit{k}_1^2 - i 0}
\int \frac{d^2 \bit{k}_2}{(2 \pi)^2}
{\not\!\! \bit{A}} (\infty, \bit{k}_2)
\frac{{\not\! \bit{k}}_1 + {\not\! \bit{k}}_2}{(\bit{k}_1 + \bit{k}_2)^2 - i 0}
\cdots
\nonumber\\
&&\qquad\qquad\times
\int \frac{d^2 \bit{k}_n}{(2 \pi)^2}
{\not\!\! \bit{A}} (\infty, \bit{k}_n)
\frac{
\sum_{i = 1}^n {\not\! \bit{k}}_i
}{
\left( \sum_{j = 1}^n \bit{k}_j \right)^2 - i 0
}
\, \gamma_\nu \psi (0)
| P \rangle \, .
\end{eqnarray}
The Fourier transformation gives
\begin{eqnarray}
\langle p_J, \bar N | j_\nu (0) | P \rangle_{(n)}
\!\!\!&=&\!\!\! (- i g)^n \bar u (p_J)
\langle \bar N |\!
\int \prod_{i = 1}^n \!\frac{d^2 \bit{\xi}_i}{2 \pi}
{\not\!\! \bit{A}} (\infty, \bit{\xi}_1)
{\not\!\! \bit{\nabla}_1} \ln | \bit{\xi}_1 - \bit{\xi}_2 |
{\not\!\! \bit{A}} (\infty, \bit{\xi}_2)
{\not\!\! \bit{\nabla}_2} \ln | \bit{\xi}_2 - \bit{\xi}_3 |
\cdots
\nonumber\\
&&\times
{\not\!\! \bit{A}} (\infty, \bit{\xi}_{n - 1})
{\not\!\! \bit{\nabla}_{n - 1}} \ln | \bit{\xi}_{n - 1} - \bit{\xi}_n |
{\not\!\! \bit{A}} (\infty, \bit{\xi}_n)
{\not\!\! \bit{\nabla}_n} \ln | \bit{\xi}_n |
\, \gamma_\nu \psi (0)
| P \rangle \, .
\end{eqnarray}
Now, exploiting Eq.\ (\ref{PureGauge}), we integrate by parts starting
with $\bit{\xi}_1$ and using ${\not\!\! \bit{\nabla}_1}^2 \ln | \bit{\xi}_1
- \bit{\xi}_2| = - \bit{\nabla}_1^2  \ln | \bit{\xi}_1 - \bit{\xi}_2|
= - 2 \pi \delta^{(2)} (\bit{\xi}_1 - \bit{\xi}_2)$, then with respect
to $\bit{\xi}_2$, first noticing that
\begin{equation}
\phi (\bit{\xi}) \bit{A}_\alpha (\bit{\xi})
=
\frac{1}{2} \bit{\nabla}_\alpha \phi^2 (\bit{\xi}) \, ,
\end{equation}
and then performing the same partial integrations as for $\bit{\xi}_1$.
In this way one can integrate out all $\bit{\xi}$'s. At the end, one
gets
\begin{equation}
\langle p_J, \bar N | j_\nu (0) | P \rangle_{(n)}
= (- i g)^n \bar u (p_J) \gamma_\nu
\langle \bar N |
\frac{1}{n !} \phi^n (\bit{0}) \psi (0)
| P \rangle \, .
\end{equation}
As a final step, one simply notices that
\begin{equation}
\frac{(-1)^n}{n !} \phi^n (\bit{0})
=
\int_0^\infty
d \bit{\xi}_1 \cdot \bit{A} (\infty, \bit{\xi}_1)
\int_0^{\bit{\scriptstyle \xi}_1}
d \bit{\xi}_2 \cdot \bit{A} (\infty, \bit{\xi}_2)
\cdots
\int_0^{\bit{\scriptstyle \xi}_{n - 1}}
d \bit{\xi}_n \cdot \bit{A} (\infty, \bit{\xi}_n)
\, ,
\end{equation}
so that it forms a gauge link once resummed to all orders
\begin{equation}
\label{DIStransverseLink}
[\infty, \bit{\infty}; \infty, \bit{0}]
=
P \exp
\left(
i g \int_0^\infty d \bit{\xi} \cdot \bit{A} (\infty, \bit{\xi})
\right) \, .
\end{equation}
Therefore, restoring the light-cone gauge link, one finds the complete
result for the amplitude
\begin{equation}
\label{FullAmplitude}
\langle p_J, \bar N | j_\nu (\xi) | P \rangle
=
\bar u (p_J) \gamma_\nu
\langle \bar N |
[\infty, \bit{\infty}; \xi_-, \bit{\xi}]_C \, \psi (\xi_-, \bit{\xi})
| P \rangle \, ,
\end{equation}
where
\begin{equation}
[\infty, \bit{\infty}; \xi_-, \bit{\xi}]_C
\equiv
[\infty, \bit{\infty}; \infty, \bit{\xi}]
[\infty, \bit{\xi}; \xi_-, \bit{\xi}] \, ,
\end{equation}
is shown in Fig.\ \ref{LinksPDFtransverse}.

Multiplying (\ref{FullAmplitude}) by its complex conjugate, we deduce the
gauge invariant transverse mo\-mentum-dependent parton distribution
\begin{equation}
\label{DIStransversePDF}
q (x, \bit{k}) = \frac{1}{2}
\int \frac{d \xi_-}{2 \pi} \frac{d^2 \bit{\xi}}{(2 \pi)^2}
{\rm e}^{- i x \xi_- + i \bit{\scriptstyle k} \cdot \bit{\scriptstyle \xi}}
\langle P |
\bar\psi (\xi, \bit{\xi}) [\infty, \bit{\infty}; \xi_-, \bit{\xi}]^\dagger_C
\, \gamma_+
[\infty, \bit{\infty}; 0, \bit{0}]_C \, \psi (0, \bit{0})
| P \rangle \, .
\end{equation}
The unitarity implies a partial cancellation of links at light-cone infinity
\begin{equation}
[\infty, \bit{\infty}; \xi_-, \bit{\xi}]^\dagger_C
[\infty, \bit{\infty}; 0, \bit{0}]_C
=
[\infty, \bit{\xi}; \xi_-, \bit{\xi}]^\dagger
[\infty, \bit{\xi}; \infty, \bit{0}]
[\infty, \bit{0}; 0, \bit{0}]
\, ,
\end{equation}
so that the definition (\ref{PDFliterature}) of the parton distribution,
accepted in the literature, acquires an additional transverse link.

\section{Parton distribution in scalar QED}
\label{PDFinQEDmodel}

In this section we demonstrate, with a model calculation of parton
distribution functions in scalar QED \cite{DucBroHoy92,BroHoyMag96},
the role played by the transverse gauge link in the restoration of a gauge
invariant answer on the amplitude level. In Ref. \cite{BroHoyMarPeiSan01},
this model was used to demonstrate the presence of non-pinched final
state interactions in scaling contribution to the deeply inelastic cross
section. We reformulate this in the language of parton distributions
and show how the final state interaction can be shifted from the final
to the initial state depending on the prescription on the spurious pole
in the gluon propagator in the light-cone gauge.

The model considered in Ref. \cite{BroHoyMarPeiSan01} consists of
heavy $D$ and light $\phi$ charged scalars with masses $M$ and $m$,
respectively, interacting with massive U(1) gauge fields $A_\mu$,
\begin{equation}
{\cal L}_{\rm sQED}
= \left( {\cal D}_\mu \phi \right)^\dagger {\cal D}_\mu \phi
+ \left( {\cal D}_\mu D \right)^\dagger {\cal D}_\mu D
- m^2 \phi^\dagger \phi - M^2 D^\dagger D
- \frac{1}{4} F_{\mu\nu}^2 + \frac{\lambda^2}{2} A_\mu^2
\, ,
\end{equation}
via the covariant derivative ${\cal D}_\mu \equiv \partial_\mu + i g A_\mu$.
Using the electromagnetic current of scalar QED
\begin{equation}
j_\mu = i \phi^\dagger
\left(
\stackrel{\rightarrow}{\partial}_\mu
-
\stackrel{\leftarrow}{\partial}_\mu
\right) \phi \, ,
\end{equation}
in the hadronic tensor (\ref{DIStensorOne}), a simple calculation gives
for the longitudinal structure function $F_L$ in the handbag
approximation\footnote{Recall that $F_1 \sim \sigma_T$ is zero for scalar
quarks to this order.}
\begin{equation}
F^{\rm tw-2}_L (\Bx, Q^2)
=
2 \Bx \left\{ \pi (\Bx) - \pi (- \Bx) \right\} \, ,
\end{equation}
with the scalar $\phi$-parton distribution in the $D$-meson
\begin{equation}
\label{PionDistribution}
\pi (x) = \frac{1}{2} \int \frac{d \xi_-}{2 \pi} {\rm e}^{i x \xi_-}
\langle P |
\phi^\dagger (0, \bit{0}) [\infty, \bit{\infty}; 0, \bit{0}]^\dagger_C \,
i \partial_+
[\infty, \bit{\infty}; \xi_-, \bit{0}]_C \, \phi (\xi_-, \bit{0})
| P \rangle \, .
\end{equation}
Here the subscript $C$ denotes the path shown in Fig.\
\ref{LinksPDFtransverse}. In practical computations we insert a complete
set of intermediate states, so that the parton distribution reads
\begin{equation}
\pi (x) = \frac{x}{2} \int \frac{d \xi_-}{2 \pi} {\rm e}^{i x \xi_-}
\sum_N \int \prod_{k = 1}^N \frac{d^4 p_k}{(2 \pi)^4}
(2 \pi) \delta_+ \left( p_k^2 - m_k^2 \right)
{\cal A}^\dagger_N (0) {\cal A}_N (\xi_-) \, ,
\end{equation}
where we introduced the amplitude
\begin{equation}
{\cal A}_N (\xi_-) \equiv
\langle p_1, \dots , p_N |
[\infty, \bit{\infty}; \xi_-, \bit{0}]_C \, \phi (\xi_-, \bit{0})
| P \rangle \, .
\end{equation}

In the following, we consider two-particle final state $N = 2$ with momenta
$p_1 = - p$ and $p_2 = P'$. The amplitude expanded in perturbation series
takes the form
\begin{equation}
{\cal A}_N (\xi_-)
= \int \frac{d^4 k}{(2 \pi)^4} {\rm e}^{i \xi_- (p - k)_+}
(2 \pi)^4 \delta^{(4)} (k + P^\prime - P)
\sum_{n = 1}^\infty g^{2 n} A^{(n)}_N (k) \, ,
\end{equation}
where $k$ is the total $t$-channel momentum
\begin{equation}
\label{TotalMomentum}
k = \sum_{i = 1}^n k_i \, ,
\end{equation}
transferred from the target. For the small-$x$ behavior, we are presently
discussing, gluon $t$-channel intermediate states dominate, so that $k_i$ is
the momentum of the $i$th gluon. Integrating out $P'_\mu$, $k_{\pm}$ and $p_+$,
we get for the small-$x$ distribution function (\ref{PionDistribution})
\begin{equation}
\pi (x) = \frac{x}{2}
\int \frac{d p_-}{p_-}
\int \frac{d^2 \bit{k}}{(2 \pi)^2} \frac{d^2 \bit{p}}{(2 \pi)^2}
\alpha_s^2
\left|
\sum_{n = 1}^\infty g^{2 (n - 1)}A^{(n)}_N (\bit{k})
\right|^2 \, .
\end{equation}
The integrated components of momenta are set to
\begin{equation}
\label{Momenta}
k_+ = \frac{D (\bit{p})}{2 p_-}
\, , \qquad
p_+ = \frac{\bit{p}^2 + m^2}{2 p_-}
\, , \qquad
k_- \approx - \bit{k}^2 \, ,
\end{equation}
by means of momentum conservation and on-mass-shell conditions. Here
we introduced the convention
\begin{equation}
D (\bit{p}) \equiv 2 x p_- + \bit{p}^2 + m^2 - i0 \, .
\end{equation}

Now we are in a position to perform the computation of the amplitude
${\cal A}_N$ in different gauges in order to demonstrate the important
role played by the transverse gauge link in generating a gauge
invariant answer.

\subsection{Feynman gauge}
\label{FeynmanGaugeComputation}

\begin{figure}[t]
\begin{center}
\mbox{
\begin{picture}(0,80)(110,0)
\put(0,1){\insertfig{7}{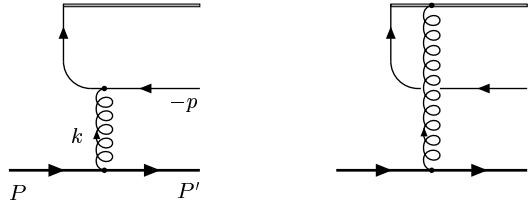}}
\end{picture}
}
\end{center}
\caption{\label{OneLoopFeynman} One-gluon exchange contribution in the
model calculation of scalar parton distribution in the Feynman gauge.}
\end{figure}

First, we turn to the Feynman gauge. One-gluon exchange, see Fig.\
\ref{OneLoopFeynman}, trivially yields
\begin{equation}
\label{FeynmanContributionOne}
A^{(1)}_2 =
\frac{
(2p - k) \cdot (2P - k)
}{
((k - p)^2 - m^2 + i 0)
(k^2 - \lambda^2 + i 0)
}
+
\frac{
(2P - k)_+
}{
(k_+ - i 0)
(k^2 - \lambda^2 + i 0)
} \, .
\end{equation}
Employing the approximation (\ref{Momenta}), one finds
\begin{equation}
\label{OneGluonExchange}
A^{(1)}_2
=
\frac{4 p_-}{( \bit{k}^2 + \lambda^2 - i0 )}
\left\{
\frac{1}{D (\bit{p} - \bit{k})} - \frac{1}{D (\bit{p})}
\right\} \, ,
\end{equation}
so that the small-$x$ parton distribution reads to this order
\begin{equation}
\label{LeadingOrderDistributionScalar}
\pi (x) = 2 x \int d p_- p_-
\int \frac{d^2 \bit{k}}{(2 \pi)^2} \frac{d^2 \bit{p}}{(2 \pi)^2}
\frac{4 \alpha_s^2}{( \bit{k}^2 + \lambda^2 - i0 )^2}
\left\{
\frac{1}{D (\bit{p} - \bit{k})} - \frac{1}{D (\bit{p})}
\right\}^2 \, .
\end{equation}

\begin{figure}[t]
\begin{center}
\mbox{
\begin{picture}(0,200)(180,0)
\put(0,1){\insertfig{12}{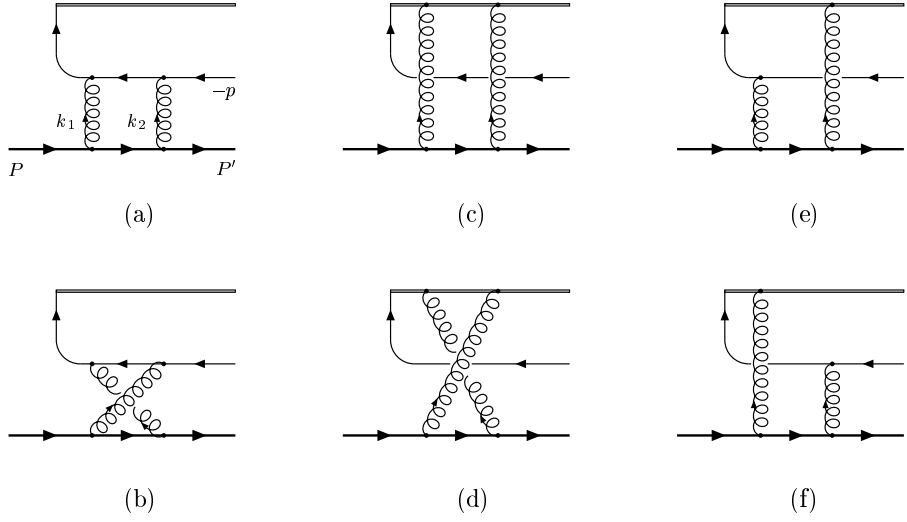}}
\end{picture}
}
\end{center}
\caption{\label{TwoLoopFeynman} One-loop diagrams in the Feynman gauge.}
\end{figure}

We now consider the one-loop contributions to the parton amplitude $A_2$.
In the small $x$ region, the dominant contribution comes, as we already
mentioned above, from diagrams with two-gluon intermediate states. There
are six such Feynman diagrams shown in Fig.\ \ref{TwoLoopFeynman}. As in
Ref.\ \cite{BroHoyMarPeiSan01}, the gauge interaction between the $D$-meson
and the rest is dominated by static Coulomb exchanges. It is easy to see
that in the small-$x$ limit,
\begin{equation}
A_{2 {\rm (a + b)}}^{(2)}
=
i g^2 2 p_-
\frac{1}{D( \bit{p} - \bit{k} )}
\int \frac{d^2 \bit{k}_2}{(2 \pi)^2}
\frac{1}{( \bit{k}_1^2 + \lambda^2 - i 0 )( \bit{k}_2^2 + \lambda^2 - i 0 )}
\, ,
\end{equation}
where we have integrated over $k_{2 -}$ and $k_{2 +}$. On the other hand,
the diagrams with two eikonal interactions generate the amplitude
\begin{equation}
A_{2 {\rm (c + d)}}^{(2)}
=
i g^2
\frac{1}{k_+ - i 0}
\int \frac{d^2 \bit{k}_2}{(2 \pi)^2}
\frac{1}{( \bit{k}_1^2 + \lambda^2 - i 0 )( \bit{k}_2^2 + \lambda^2 - i 0 )}
\, .
\end{equation}
Finally, we include the contribution for diagrams \ref{TwoLoopFeynman}
(e) and (f)
\begin{equation}
A_{2 {\rm (e + f)}}^{(2)}
=
- i g^2 2 p_-
\int \frac{d^2 \bit{k}_2}{(2 \pi)^2}
\frac{1}{( \bit{k}_1^2 + \lambda^2 - i 0 )( \bit{k}_2^2 + \lambda^2 - i 0 )}
\left\{
\frac{1}{D ( \bit{p} - \bit{k}_1 )}
+
\frac{1}{D ( \bit{p} - \bit{k}_2 )}
\right\}
\, .
\end{equation}
where the minus sign indicates that the interference term is destructive.
Summing over all contributions and using Eq.\ (\ref{Momenta}), we have the
one-loop contribution
to the parton amplitude,
\begin{eqnarray}
\label{OneLoop}
A^{(2)}_2
\!\!\!&=&\!\!\!
i \frac{g^2}{2!} 4 p_-
\int \frac{d^2 \bit{k}_2}{(2 \pi)^2}
\frac{1}{( \bit{k}_1^2 + \lambda^2 - i 0 )( \bit{k}_2^2 + \lambda^2 - i 0 )}
\nonumber \\
&&\qquad\qquad\qquad\times
\left\{
\frac{1}{D ( \bit{p} - \bit{k} )}
-
\frac{1}{D ( \bit{p} - \bit{k}_1 )}
-
\frac{1}{D ( \bit{p} - \bit{k}_2 )}
+
\frac{1}{D ( \bit{p} )}
\right\}
\, ,
\end{eqnarray}
where $\bit{k} = \bit{k}_1 + \bit{k}_2$ according to (\ref{TotalMomentum}).
When the $\bit{p} \to \infty$ the amplitude vanishes as $1/\bit{p}^4$ or
as $\bit{r}^2$ with $\bit{r} \to 0$ when Fourier transformed. The argument
is straightforward: The Wilson line represents an infinite-energy particle
which has the opposite charge as compared to the spectator. When the distance
between them goes to zero the gluon does not resolve this neutral composite
system and the interaction vanishes. Eq.\ (\ref{OneLoop}) agrees with the
result of Ref.\ \cite{BroHoyMarPeiSan01}.

Eqs.\ (\ref{OneGluonExchange}) and (\ref{OneLoop}) are the first two terms
in the expansion of the exponent. A computation of higher orders leads to the
result for the amplitude
\begin{eqnarray}
\label{ExponentiatedAmplitude}
{\cal A}_2 (0)
\!\!\!&=&\!\!\! i 4 p_- \int d^2 \bit{r} \,
{\rm e}^{i \bit{\scriptstyle p} \cdot \bit{\scriptstyle r}}
\,
\frac{1}{2 \pi} K_0 \left( \bit{r} \sqrt{2 \Bx p_- + m^2} \right)
\nonumber\\
&&\times
\int d^2 \bit{R} \,
{\rm e}^{i (\bit{\scriptstyle P} - \bit{\scriptstyle P}')
\cdot
\bit{\scriptstyle R}}
\,
\bigg\{
1
-
\exp
2 i \alpha_s
\bigg(
K_0 \left( ( \bit{R} + \bit{r} ) \lambda \right)
-
K_0 \left( \bit{R} \lambda \right)
\bigg)
\bigg\} \, .
\end{eqnarray}
It has a factorized form of a convolution of the photon wave function
$K_0 \left( \bit{r} \sqrt{2 \Bx p_- + m^2} \right)$ and the scattering
amplitude of the $\phi \phi^\ast$-pair, separated by the distance $\bit{r}$,
of the struck quark, represented by the path-ordered exponential, and the
spectator off the $D$-meson. In section \ref{Dipoles}, we reproduce this
equation in the aligned-jet approximation for the dipole picture of
$\gamma^\ast D$ scattering.

\subsection{Light-cone gauge}
\label{LightConeGaugeComputations}

\begin{figure}[t]
\begin{center}
\mbox{
\begin{picture}(0,115)(110,0)
\put(0,1){\insertfig{7}{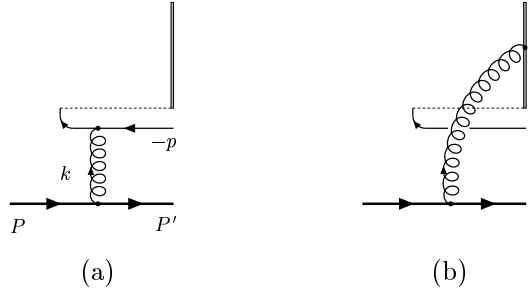}}
\end{picture}
}
\end{center}
\caption{\label{OneLoopLightCone} One-gluon diagrams for the parton
distribution in the light-cone gauge. The dashed line stands for absent
light-cone gauge like and serves merely the purpose to guide the eye.}
\end{figure}

In this section, we calculate the same parton distribution in the
light-cone gauge $A_+ = 0$. As we already emphasized in the introduction,
this condition does not fix the gauge completely since the $x_-$-independent
gauge transformations leave it invariant. As a result the perturbation
theory is not defined due to ambiguities in contour deformation of momentum
integrals due to presence of an $1/k_+$-singularity in
the gauge propagator
\begin{equation}
\label{PropagatorLCunreg}
D_{\mu\nu} (k) = \frac{1}{k^2 + i 0}
\left(
g_{\mu\nu}
-
\frac{k_\mu n_\nu + k_\nu n_\mu}{k_+}
\right) \, .
\end{equation}
Most attempts in the literature to address the problem were pragmatic. A
prescription was invented and then tested in practical loop computations.
A justification by rigorous quantization procedures came later if ever
\cite{BasNarSol91,Lei94}. The fixing of residual gauge degrees of freedom,
or maximal gauge fixing, can be done by means of boundary conditions on
the gauge potential as is spelled out in the appendix \ref{Residual}. With
this procedure one gets a set of regularization prescriptions on the
light-cone pole corresponding to advanced, retarded and antisymmetric
boundary conditions for the gluon potential,
\begin{equation}
\frac{1}{[k_+]} =
\left\{
\begin{array}{ll}
{\rm Adv:} & \frac{1}{k_+ - i 0} \\
{\rm Ret:} & \frac{1}{k_+ + i 0} \\
{\rm PV:}  & \frac{1}{2} \left\{ \frac{1}{k_+ + i 0} + \frac{1}{k_+ - i 0} \right\}
\end{array}
\right.
\, .
\end{equation}
Here only the pole associated with $k_\nu n_\mu$ term, where the momentum
$k$ enters the vertex $\nu$, of the gluon density matrix is displayed
since only this part works in calculations performed later in this section.
The other contribution can be obtained by hemiticity.

In principle, any prescription for the light-cone singularity should work,
so long as it is easily generalizable to multi-loop calculations. Since all
of the above prescriptions are not causal, loop integrations in particular
Feynman diagrams can be pinched between poles and result into regularized
spurious divergences. So that off-shell Green functions and renormalization
constants do depend on the prescription at hand and, therefore, contain
light-cone singularities. However, for a gauge invariant set of diagrams
they cancel as they should, see, e.g., \cite{CurFurPet80} for a two-loop
computation with the principal value prescription.

In our specific computation, the amplitude ${\cal A}_N$, with the transverse
gauge link, is gauge invariant by itself and independent on the regularization
procedure. However, it is obvious that in the light-cone distribution
(\ref{LightConeDistribution}) the extra link does not contribute in the product
of the parton amplitude and its complex conjugate due to the unitarity
condition $[\infty, \bit{\infty}; 0, \bit{\infty}]^\dagger [\infty, \bit{\infty};
0, \bit{\infty}] = 1$. In the following subsections, we verify these statements
explicitly through the example of the previous section.

\begin{figure}[t]
\begin{center}
\mbox{
\begin{picture}(0,209)(165,0)
\put(0,1){\insertfig{11}{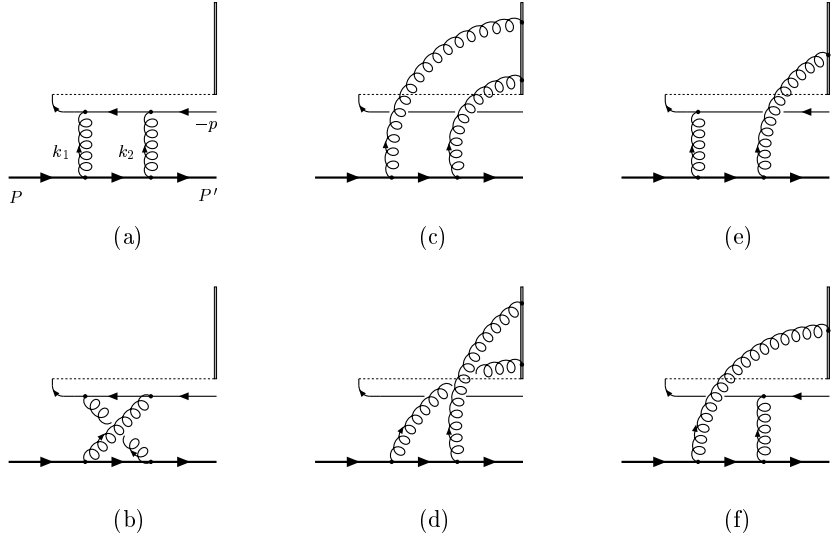}}
\end{picture}
}
\end{center}
\caption{\label{TwoLoopLightCone} One-loop diagrams in the light-cone gauge.}
\end{figure}

To start with, let us address the tree level diagrams first. The contribution
from the first term in the gluon propagator is the result of Fig.\
\ref{OneLoopFeynman} (a) in the Feynman gauge. The term $k_\mu n_\nu$ yields
a null contribution because $k_\mu$ contracted with the scalar $D$-meson vertex
$(P + P')_\mu$ vanishes by the on-shell condition. Contribution from Fig.\
\ref{OneLoopLightCone} (a) is thus
\begin{equation}
\label{TreeLCspectator}
A_{2 {\rm (a)}}^{(1)}
= \frac{1}{((k - p)^2 - m^2 + i 0) (k^2 - \lambda^2 + i 0)}
\left\{
(2p - k) \cdot (2 P - k)
- (2 P - k)_+ k \cdot (2 p - k) \frac{1}{[k_+]}
\right\} \, .
\end{equation}
Next we exploit the simple fact
\begin{equation}
k \cdot (2 p - k) = (p^2 - m^2) - ((k - p)^2- m ^2) \, ,
\end{equation}
where the first term dies out by the on-shell condition, in order to cancel
the pion propagator in the amplitude with the second term. In the expression
for the transverse gauge-link contribution we encounter a factor handled as
\begin{equation}
\frac{{\rm e}^{- i \infty k_+}}{[k_+]} = 2 \pi i \chi \delta (k_+) \, ,
\end{equation}
with $\chi$ depending on the prescription adopted for the pole
\begin{equation}
\label{ChiValues}
\chi
=
\left\{
\begin{array}{ll}
{\rm Adv:} & \phantom{-} 0 \\
{\rm Ret:} & - 1 \\
{\rm PV:}  & - \frac{1}{2}
\end{array}
\right.
\, .
\end{equation}
As we expected, the advanced boundary condition, which sets the transverse link
to unity, is supported by the perturbative calculation. Adding the contribution
of the extra gauge link, Fig.\ \ref{OneLoopLightCone} (b),
\begin{equation}
A_{2 {\rm (b)}}^{(1)}
=
- \frac{(2 P - k)_+}{k^2 - \lambda^2 + i 0} 2 \pi i \chi \delta (k_+)
\, ,
\end{equation}
to (\ref{TreeLCspectator}), one gets the result of the Feynman gauge,
Eq.\ (\ref{FeynmanContributionOne}).

Let us note in advance that in multi-gluon exchange diagrams, similarly
to the one-gluon case, only $k_\nu n_\mu$ part of the gluon density
matrix gives the leading contribution. The reason is that $k_\mu n_\nu$
part when contracted with the heavy meson line yields
\begin{equation}
k_i \cdot \left( 2 \left( P - \sum_{j = 1}^{i - 1} k_j \right) - k_i \right)
=
\left\{ \left( P - \sum_{j = 1}^{i - 1} k_j \right)^2 - M^2 \right\}
-
\left\{ \left( P - \sum_{j = 1}^{i} k_j \right)^2 - M^2 \right\}
\, ,
\end{equation}
and since the dominant effect comes from the on-mass-shell $D$-meson
intermediate states, both terms vanish independently.

\subsubsection{Advanced prescription}

Since the advanced prescription is implied by the advanced boundary
condition $\bit{A} (\infty) = 0$, see the Appendix \ref{Residual},
the transverse gauge link at light-cone infinity does not contribute.
This was observed in Ref.\ \cite{JiYua02} in a computation of a
transverse momentum-dependent parton distribution.

In this section, we perform an explicit calculation to reproduce the
result of section \ref{FeynmanGaugeComputation} in the light-cone gauge.
Instead of showing the correspondence on a diagram-by-diagram basis in
different gauges, we present the result of calculations for subsets of
graphs. From diagrams (a) and (b) in Fig.\ \ref{TwoLoopLightCone}, we obtain
\begin{eqnarray}
A_{\rm (a + b)}
\!\!\!&=&\!\!\! i g^2 4 p_-
\int \frac{d^2 \bit{k}_2}{(2\pi)^2}
\frac{1}{( \bit{k}_1^2 + \lambda^2 - i 0 )( \bit{k}_2^2 + \lambda^2 - i 0 )}
\nonumber\\
&&\qquad\qquad\qquad\qquad\qquad\qquad\times\frac{
\left(
D(\bit{p} - \bit{k}_2 ) - D( \bit{p} - \bit{k})
\right)
\left(
D(\bit{p}) - D (\bit{p} - \bit{k}_2)
\right)
}{
D (\bit{p}) D (\bit{p} - \bit{k}) D (\bit{p} - \bit{k}_2)
} \, .
\end{eqnarray}
However, this is not the complete answer. In the present case, we have to
take into account an additional diagram shown in Fig.\ \ref{ExtraLightCone}.
It is negligible in the Feynman gauge, but not in the light-cone gauge. The
contribution of this graph is
\begin{equation}
A_{\rm ex}
=
i g^2 2 p_-
\int \frac{d^2 \bit{k}_2}{(2\pi)^2}
\frac{1}{( \bit{k}_1^2 + \lambda^2 - i 0 )( \bit{k}_2^2 + \lambda^2 - i 0 )}
\frac{
D( \bit{p} - \bit{k}_1 )
+
D( \bit{p} - \bit{k}_2 )
-
D( \bit{p} )
-
D( \bit{p} - \bit{k} )
}{
D( \bit{p} ) D( \bit{p} - \bit{k} )
} \, .
\end{equation}
The sum of the above two equations is exactly the right-hand side of Eq.\
(\ref{OneLoop}) apart from terms which integrate to zero by the symmetry
property of the integrand.

The physics of the advanced prescription is that all final state interactions
have been shifted to the initial state light-cone wave functions. Parton
distributions are truly densities of partons. This scheme is the closest
one to the original formulation of the Feynman parton model.

\subsubsection{Principal value prescription}

\begin{figure}[t]
\begin{center}
\mbox{
\begin{picture}(0,80)(55,0)
\put(0,1){\insertfig{3}{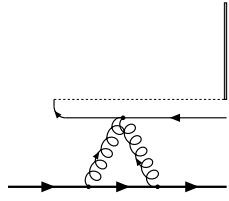}}
\end{picture}
}
\end{center}
\caption{\label{ExtraLightCone} One-loop diagram negligible in the Feynman
gauge but contributing in the light-cone gauge.}
\end{figure}

Now we discuss the principal value regularization for the light-cone
singularity. One virtue of this prescription is that it does not
introduce an absorptive part in the light-cone wave functions. Final state
scattering phases then appear only through the extra gauge link derived in
section \ref{TransverseGaugeLink}, as have been shown in the example of
the single transverse-spin asymmetry in semi-inclusive deeply inelastic
scattering \cite{JiYua02}.

A straightforward calculation gives the following expression for Fig.\
\ref{TwoLoopLightCone} (a) and (b),
\begin{equation}
A_{\rm (a + b)}
= i g^2 2 p_-
\int \frac{d^2 \bit{k}_2}{(2\pi)^2}
\frac{1}{( \bit{k}_1^2 + \lambda^2 - i 0 )(  \bit{k}_2^2 + \lambda^2 - i 0 )}
\left\{
\frac{1}{D( \bit{p} - \bit{k})}
-
\frac{1}{D(\bit{p} - \bit{k}_2)}
\right\} \, .
\end{equation}
The diagram in Fig.\ \ref{ExtraLightCone} does not contribute. The
transverse gauge link generates the diagrams (c), (d), (e) and (f)
in Fig.\ \ref{TwoLoopLightCone}, whose straightforward computation
gives,
\begin{equation}
A_{\rm (c + d + e + f)}
=
i g^2 2 p_-
\int \frac{d^2 \bit{k}_2}{(2\pi)^2}
\frac{1}{( \bit{k}_1^2 + \lambda^2 - i 0 )( \bit{k}_2^2 + \lambda^2 - i 0 )}
\left\{
\frac{1}{D( \bit{p} )} - \frac{1}{D( \bit{p} - \bit{k}_1 )}
\right\} \, .
\label{TransvLinkContr}
\end{equation}
Again the sum of the two contributions yields the same result as in
Eq.\ (\ref{OneLoop}).

The gauge link contribution in the parton distribution is unitary. When
integrated over $\bit{k}$, it does not contribute. Therefore, the square
of the amplitude in Eq.\ (\ref{TransvLinkContr}) must cancel against
an interference between the two-loop gauge-link diagrams and the tree.
This is indeed what has been observed in Ref.\ \cite{BroHoyMarPeiSan01}.
The present discussion demystifies this cancellation.

To conclude, the new gauge link found in section \ref{TransverseGaugeLink}
is necessary for reproducing the Feynman-gauge parton amplitude in the
light-cone gauge calculation. However, it does not contribute to the
integrated parton distribution. In particular, it does not play a role
in modeling of leading twist ``shadowing''.

\subsubsection{Phase due to transverse link}

Actually, using the exponentiation property of soft radiation, one
can compute the phase generated by the transverse link exactly in the
limit of infinitely heavy mass of the $D$-meson, $M \to \infty$. In
this situation one can approximate the $D$-meson propagation by an eikonal
line along its four-velocity $v_\mu \equiv P_\mu/M$. The parton amplitude
reads
\begin{equation}
{\cal A}_2
\stackrel{M \to \infty}{\longrightarrow}
\int d^2 \bit{R} \,
{\rm e}^{i (\bit{\scriptstyle P} - \bit{\scriptstyle P}')
\cdot
\bit{\scriptstyle R}}
\langle p |
T \left\{
[\infty v, \bit{R}; - \infty v, \bit{R} ]
[\infty, \bit{\infty}; \infty, \bit{0}]
\phi (0, \bit{0})
\right\}
| 0 \rangle
\, .
\end{equation}
Extracting the interaction of the $\phi$-spectator with the target, we get
\begin{equation}
{\cal A}_2
\approx
\int d^2 \bit{R} \,
{\rm e}^{i (\bit{\scriptstyle P} - \bit{\scriptstyle P}')
\cdot
\bit{\scriptstyle R}}
a (\bit{R} \lambda) W (\bit{R} \lambda) \, .
\end{equation}
where $a$ stands for the $\phi D$-interaction as well as self-interactions, while
\begin{equation}
W (\bit{R} \lambda)
=
\frac{\langle 0 |
T \left\{
[\infty v, \bit{R}; - \infty v, \bit{R} ]
[\infty, \bit{\infty}; \infty, \bit{0}]
\right\}
| 0 \rangle
}{
\langle 0 |
T \left\{
[\infty v, \bit{R}; - \infty v, \bit{R} ]
\right\}
| 0 \rangle
\langle 0 |
T \left\{
[\infty, \bit{\infty}; \infty, \bit{0}]
\right\}
| 0 \rangle
} \, ,
\end{equation}
represents the very effect of the transverse gauge link on the amplitude.
In an abelian theory, the correlator of Wilson lines exponentiates exactly,
\begin{equation}
\label{Exp1}
W (\bit{R} \lambda)
=
{\rm e}^{i \varphi (\bit{\scriptstyle R} \lambda)} \, ,
\end{equation}
where
\begin{equation}
\label{Exp2}
\varphi = g^2
\int_{C_1} d x_\mu \int_{C_2} d y_\nu
D_{\mu\nu} (x - y) \, ,
\end{equation}
is given in terms of the gluon propagator $\langle 0 | T \left\{ A_\mu (x)
A_\nu (y) \right\} | 0 \rangle \equiv - i D_{\mu\nu} (x - y)$ and the integration
goes over the two pieces of path $C_1 = [\infty v, \bit{R}; - \infty v, \bit{R} ]$
and $C_2 = [\infty, \bit{\infty}; \infty, \bit{0}]$.
A simple computation yields
\begin{equation}
\varphi (\bit{R} \lambda)
=
\chi g^2
\int \frac{d^2 \bit{k}}{(2 \pi)^2}
\frac{
{\rm e}^{- i \bit{\scriptstyle k} \cdot {\bit{\scriptstyle R}}}
}{
\bit{k}^2 + \lambda^2 - i 0} \, ,
\end{equation}
with
\begin{equation}
\chi \equiv \int \frac{d k_+}{2 \pi i} \frac{{\rm e}^{- i \infty k_+}}{[k_+]} \, .
\end{equation}
In the propagator (\ref{PropagatorLCunreg}) only the term $k_\nu n_\mu$, where
$k$ flows into the vertex of the transverse gauge link, works. Using Eq.\
(\ref{ChiValues}), one finds
\begin{equation}
\label{Phases}
\varphi_{\rm Adv} (\bit{R} \lambda) = 0
\, ,
\qquad
\varphi_{\rm PV} (\bit{R} \lambda) = - \alpha_s K_0 (\bit{R} \lambda) \, .
\end{equation}
This is an exact all-order result in the abelian theory, we are considering, and
agrees with the lowest order predictions of Ref.\ \cite{BroHoyMarPeiSan01}. In
a non-abelian theory, $\varphi$ admits the web expansion \cite{Ste81,Gat83,FreTay84}
\begin{eqnarray*}
\varphi = \sum_w C_w G_w \, ,
\end{eqnarray*}
where $C_w$ is the maximally non-abelian color factor and $G_w$ is a sum of
corresponding web graphs. Eq.\ (\ref{Phases}) is the first term in the web
expansion.

\section{Dipole picture}
\label{Dipoles}

This section is devoted to the demonstration of the equivalence of
twist-two approximation for the structure functions at small values of
the Bjorken variable and the aligned-jet kinematics of the dipole
description of the $\gamma^\ast$-hadron scattering. In the latter formalism,
at small-$\Bx$, the virtual photon splits into a $q \bar q$-pair long
before the interaction with the target, due to large Ioffe time $\tau
\sim 1/(M \Bx)$. On the second stage, the dipole interacts with the hadron.

First we show this correspondence with the lowest order calculation
in the scalar model alluded to above and then turn to an all-order proof
in the semiclassical approach to small-$\Bx$ QCD processes.

\subsection{Dipole scattering in scalar QED}

\begin{figure}[t]
\begin{center}
\mbox{
\begin{picture}(0,100)(70,0)
\put(0,1){\insertfig{5}{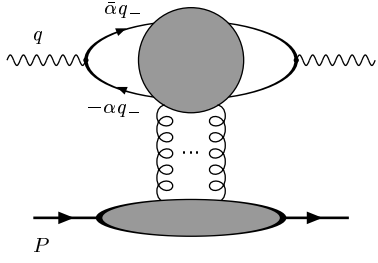}}
\end{picture}
}
\end{center}
\caption{\label{Dipole} Dipole scattering off hadron target.}
\end{figure}

A simple calculation of the diagram in Fig.\ \ref{Dipole} within the scalar
model discussed above produces
\begin{equation}
F^{\rm dipole}_L (\Bx, Q^2)
= Q^4 \int_0^1 d \alpha \, \alpha \bar\alpha (1 - 2 \alpha)^2
\int \frac{d^2 \bit{k}}{(2 \pi)^2} \frac{d^2 \bit{p}}{(2 \pi)^2}
\frac{4 \alpha_s^2}{( \bit{k}^2 + \lambda^2 - i0 )^2}
\left\{
\frac{1}{D (\bit{p} - \bit{k})} - \frac{1}{D (\bit{p})}
\right\}^2 \, ,
\end{equation}
where
\begin{equation}
D (\bit{k}) \equiv \alpha \bar\alpha Q^2 + \bit{k}^2 + m^2 - i 0 \, ,
\end{equation}
with $\bar\alpha \equiv 1 - \alpha$ and momentum variables having the same
meaning as in Eq.\ (\ref{LeadingOrderDistributionScalar}). Its Fourier
transformation into the coordinate space gives
\begin{equation}
\label{DipoleScalar}
F^{\rm dipole}_L (\Bx, Q^2)
= \frac{Q^4}{(2 \pi)^4}
\int_0^1 d \alpha \, \alpha \bar\alpha (1 - 2 \alpha)^2
\int d^2 \bit{r} K_0^2 \left( \bit{r} \sqrt{ \alpha \bar\alpha Q^2 + m^2} \right)
\sigma (\bit{r}) \, ,
\end{equation}
where the dipole cross section in two-gluon approximation reads
\begin{equation}
\sigma (\bit{r})
=
4 \alpha_s^2 \int d^2 \bit{R}
\bigg\{
K_0 \left( (\bit{R} + \bit{r}) \lambda \right)
-
K_0 \left( \bit{R} \lambda \right)
\bigg\}^2
+ {\cal O} (\alpha_s^3) \, .
\end{equation}
Computation of the further terms leads to the exponentiation of the
one-gluon exchanges similarly to Eqs.\ (\ref{Exp1}) and (\ref{Exp2}).
The dipole cross section can be expressed as a correlation function
of two Wilson lines over the hadron states,
\begin{equation}
\label{TotalTranslationDipoleXsection}
\int d^2 \bit{R} \,
\langle P'|
\left( 1 - U ( \bit{R} + \bit{r} ) U^\dagger (\bit{R}) \right)
| P \rangle
=
2 \pi \delta (\Delta_+) (2 \pi)^2 \delta^{(2)} ( \bit{\Delta} )
\sigma (\bit{r}) \, ,
\end{equation}
where $\Delta = P' - P$. This makes the above observation on the
exponentiation transparent. Here $U$ is the abelian Wilson line
\begin{equation}
\label{AbelianWilsonLine}
U(\bit{x})
\equiv
[\infty, \bit{x}; - \infty, \bit{x}]
=
\exp\left(
- i g \int_{- \infty}^{\infty} d x_- A_+ \left( x_-, \bit{x} \right)
\right) \, .
\end{equation}

There are three regions in the dipole formula for the structure function
(\ref{DipoleScalar}) which lead to scaling. The small-$\bit{r}$ region,
$\bit{r}^2 \sim 1/Q^2$, corresponds to the genuine gluon contribution to the
$D$-meson structure function. The aligned-jet regions \cite{BjoKog73}:
\begin{equation}
\label{AlignedJetKin}
\alpha, \ \bar\alpha \sim \frac{{\mit \Lambda}^2}{Q^2}
\end{equation}
is where the (anti)quark component dominates. So, to extract the leading
twist-two quark contribution, we integrate in the vicinity of the end-point
$\alpha = 0$. Defining
\begin{equation}
\label{Redefinition}
\alpha
\equiv \frac{2 \Bx p_-}{Q^2} \, ,
\end{equation}
we get
\begin{equation}
F_L^{\rm al-jet} (\Bx, Q^2)
= 4 \frac{\Bx^2}{(2 \pi)^4} \int d p_- p_-
\int d^2 \bit{r} K_0^2 \left( \bit{r} \sqrt{2 \Bx p_- + m^2} \right)
\sigma (\bit{r}) \, .
\end{equation}
This result scales as a function of $Q^2$ and thus defines the twist-two
parton distribution,
\begin{equation}
F_L^{\rm tw-2} (\Bx, Q^2) = F_L^{\rm al-jet} (\Bx, Q^2) \, .
\end{equation}

\subsection{Semiclassical approach}

In the previous section we have demonstrated in a model calculation that
the dipole representation for the deeply inelastic structure functions
in the aligned-jet kinematics is equivalent to the twist-two parton
distribution functions at small momentum fractions. Here we generalize
this result to all orders using the semiclassical picture of small-$\Bx$
scattering in QCD.

For the transversely polarized photon, the structure function has the
form of a convolution of the photon wave function squared and dipole
propagation though the target  \cite{NikZak90},
\begin{equation}
F_T^{\rm dipole} (\Bx, Q^2)
= 4 N_c \frac{Q^4}{(2 \pi)^4}
\int_0^1 d \alpha \, \alpha \bar\alpha \left( \alpha^2 + \bar\alpha^2 \right)
\int d^2 \bit{r} \,
K_1^2 \left( \bit{r} \sqrt{ \alpha \bar\alpha Q^2} \right)
\sigma (\bit{r}) \, .
\end{equation}
The dipole cross section, after extraction of total translations, cf.\ Eq.\
(\ref{TotalTranslationDipoleXsection}), is determined by the correlation of
Wilson line, which represent the fast-moving quarks through the gluon field
of the nucleon
\begin{equation}
\sigma (\bit{r})
=
\int d^2 \bit{R} \,
\langle\langle
\frac{1}{N_c} {\rm tr}
\left( 1 - U ( \bit{R} + \bit{r} ) U^\dagger (\bit{R}) \right)
\rangle\rangle
\, ,
\end{equation}
with $U$ being the non-abelian analogue of Eq.\ (\ref{AbelianWilsonLine}) with
matrix valued gauge field and path ordering operation. At very high energies
the color field is localized on the light cone with a profile in the
transverse space. Namely, in the Feynman gauge,
\begin{equation}
\label{ShockWave}
A_+ (z) = \delta (z_-) \beta (\bit{z})
\, , \qquad
A_- (z) = \bit{A} (z) = 0
\, ,
\end{equation}
it has the form of a shock wave \cite{Bal95,BalBel01}, with the only non-vanishing
components of the field-strength tensor $G_{+ \perp}$. A transformation to the
light-cone gauge $A_+ = 0$ with retarded boundary condition generates the pure-gauge
color field of the McLerran-Venugopalan model \cite{MclVen93},
\begin{eqnarray*}
\bit{A} (z) = \theta (z_-) \bit{\nabla} \beta (\bit{z})
\, , \qquad
A_- (z) = 0
\, .
\end{eqnarray*}
In the aligned-jet kinematics (\ref{AlignedJetKin}), the structure function
$F_2 = F_T + F_L$ can be written as
\begin{equation}
\label{AlignedJetSpinorQuarks}
F_2^{\rm al-jet} (\Bx, Q^2)
= 16 N_c \frac{\Bx^2}{(2 \pi)^4}
\int d p_- \, p_-
\int d^2 \bit{r} \,
K_1^2 \left( \bit{r} \sqrt{ 2 \Bx p_- } \right) \sigma (\bit{r}) \, .
\end{equation}
where we used the definition (\ref{Redefinition}) and took into account
that the longitudinal structure function is suppressed in this region,
$F^{\rm al-jet}_L \sim 1/Q^2$.

Now we are in a position to reproduce the same result
(\ref{AlignedJetSpinorQuarks}) computing the leading order and twist
contribution to the same structure function,
\begin{equation}
F^{\rm tw-2}_2 (\Bx, Q^2) = \Bx \left( q (\Bx) - q (- \Bx) \right) \, ,
\end{equation}
at small $\Bx$, which is given in terms of the twist-two parton distribution
\begin{equation}
\label{InegratedQuarkDistribution}
q (x) = \frac{1}{2} \int \frac{d \xi_-}{2 \pi} {\rm e}^{i x \xi_-}
\langle P |
\bar \psi (0, \bit{0})
[0, \bit{0}; \xi_-, \bit{0}]
\gamma_+
\psi (\xi_-, \bit{0})
| P \rangle \, ,
\end{equation}
see Refs.\ \cite{MclVen93,HebWei98,MclVen98} for related considerations.
Anticipating infinities due to unrestricted integrations, we introduce a
regularized quark distribution
\begin{equation}
\label{TildeQ}
\widetilde q (x)
\equiv
\frac{1}{2}
\int \frac{d \xi_-}{2 \pi} \int d \eta_- d^2 \bit{\eta} \,
{\rm e}^{i x \xi_-}
\langle P' |
\bar\psi (\eta_-, \bit{\eta})
[\eta_-, \bit{\eta}; \xi_- + \eta_-, \bit{\eta}]
\gamma_+
\psi (\xi_- + \eta_-, \bit{\eta})
| P \rangle \, ,
\end{equation}
which is related to the conventional one (\ref{InegratedQuarkDistribution}) by
\begin{equation}
\label{QtoQtilde}
\widetilde q (x)
=
2 \pi \delta (\Delta_+) (2 \pi)^2 \delta^{(2)} (\bit{\Delta})
q (x) \, ,
\end{equation}
where one sets $P' = P$ in $q (x)$.

At very small $x$ the bulk of contribution comes from the gluon component
of the nucleon wave function while the constituent quark component is
negligible. Therefore, we can approximate $q (x)$ by computing the quark
propagator in the external shock-wave field of the nucleon,
\begin{equation}
\widetilde q (x)
= - \frac{1}{2}
\int \frac{d \xi_-}{2 \pi} \int d \eta_- d^2 \bit{\eta} \,
{\rm e}^{i x \xi_-}
\langle P' | {\rm Tr}
[\eta_-, \bit{\eta}; \xi_- + \eta_-, \bit{\eta}]
\gamma_+
i S (\xi_- + \eta_-, \bit{\eta}; \eta_-, \bit{\eta})
| P \rangle \, ,
\end{equation}
where the trace is taken with respect to color and Dirac indices. The exact
quark propagator in the shock-wave background (\ref{ShockWave}) reads
\cite{Bal95,HebWei98,KovMilWei00,FerIanLeoMcl01,BalBel01}:
\begin{eqnarray}
\label{ShockWavePropagator}
i S (x; y)
\!\!\!&=&\!\!\!
i S_0 (x - y) \theta (x_- y_-)
\\
&+&\!\!\!
\int d^4 z \, \delta (z_-)
i S_0 (x - z) \gamma_-
\left\{
\theta (x_-) \theta (- y_-) U (\bit{z})
-
\theta (- x_-) \theta (y_-) U^\dagger (\bit{z})
\right\}
i S_0 (z - y)
\, , \nonumber
\end{eqnarray}
where $S_0$ is a Fourier transform of the free quark propagator
(\ref{FreeFermionPropMomentum}),
\begin{eqnarray*}
S_0 (x - y)
=
\int \frac{d^4 k}{(2 \pi)^4} {\rm e}^{- i k \cdot (x - y)} S_0 (k)
\, .
\end{eqnarray*}
Substitution of (\ref{ShockWavePropagator}) into Eq.\ (\ref{TildeQ})
and subtraction of the free propagation yields
\begin{equation}
\widetilde q (x)
= 16 N_c \frac{x}{(2 \pi)^4}
\int d p_- \, p_- \int d^2 \bit{r} \,
K_1^2 \left( \bit{r} \sqrt{2x p_-} \right)
\int d^2 \bit{R} \,
\langle P' |
\frac{1}{N_c} {\rm tr}
\left(
1 - U (\bit{R} + \bit{r}) U^\dagger (\bit{R})
\right)
| P \rangle \, .
\end{equation}
After extraction of total translations on both sides with Eqs.\
(\ref{TotalTranslationDipoleXsection}) and (\ref{QtoQtilde}), we reproduce
the result of the aligned-jet kinematics (\ref{AlignedJetSpinorQuarks}).

\section{DY versus DIS: universality of parton distributions}
\label{DYversusDIS}

In the Drell-Yan process, the gauge link in the parton distributions
arises from the initial state interactions rather than from the final
state. Using a procedure similar to the one carried out in section
\ref{DISpdfs}, we can show that the transverse momentum-dependent parton
distributions probed in DY production are,
\begin{equation}
\label{DYtransversePDF}
q_{\rm DY} (x, \bit{k})
= \frac{1}{2} \int \frac{d \xi_-}{2 \pi} \frac{d^2 \bit{\xi}}{(2 \pi)^2}
{\rm e}^{- i x \xi_- + i \bit{\scriptstyle k} \cdot \bit{\scriptstyle \xi}}
\langle P | \bar\psi (\xi_-, \bit{\xi})
[- \infty, \bit{\infty}; \xi_-, \bit{\xi}]^\dagger_{C'} \,
\gamma_+
[- \infty, \bit{\infty}; 0, \bit{0}]_{C'} \,
\psi(0, \bit{0})
| P \rangle \, ,
\end{equation}
with the gauge link
\begin{equation}
[- \infty, \bit{\infty}; \xi_-, \bit{\xi}]_{C'}
=
[- \infty, \bit{\infty}; - \infty, \bit{\xi}]
[- \infty, \bit{\xi}; \xi_-, \bit{\xi}]
\, .
\end{equation}
The light-cone infinity $\xi_- = \infty$ has been replaced by $\xi_- = -
\infty$, reflecting the fact that the annihilation partner (antiquark or
quark) from the other hadron starts its trajectory at $\xi_- = - \infty$.
As in deeply inelastic process we find again an extra contribution due to
transverse components of the gauge potential which was not accounted for
in recent treatments, see, e.g., Ref.\ \cite{BoeMul99}.

In the light-cone gauge with the principal value prescription, the light-cone
wave function is real. Therefore, the imaginary part in the parton
distributions (\ref{DIStransversePDF}) and (\ref{DYtransversePDF}) arises
from the interaction of the spectator with the gluonic field in the transverse
link. Since in the DIS distribution the transverse link enters at $\xi_-
= \infty$ while in DY at $\xi_- = - \infty$ and the gauge potentials
obey the antisymmetric boundary condition $\bit{A} (- \infty) = - \bit{A}
(\infty)$, we find that DY link differs by the sign in the phase from the
DIS link, cf.\ Eq.\ (\ref{DIStransverseLink}),
\begin{equation}
\label{DYtransverseLinkPV}
[- \infty, \bit{\infty}; - \infty, \bit{0}]_{\rm PV}
=
P \exp
\left(
- i g \int_0^\infty d \bit{\xi} \cdot \bit{A} (\infty, \bit{\xi})
\right) \, .
\end{equation}
Therefore, one can immediately understand in the lowest order calculation
the change of the overall sign of the single transverse-spin asymmetry
from DIS to DY case emphasized in Refs.\ \cite{Col02,BroHwaSch02b}. The phase
of the transverse link, which is a genuine source of the imaginary part in
the distribution for the principal value regularization of the light-cone
gauge when is interacts with the target spectators, is opposite for DIS
and DY transverse momentum-dependent parton distributions, see Eqs.\
(\ref{DIStransverseLink}) and (\ref{DYtransverseLinkPV}). This reflects a
breakdown of the naive universality of unintegrated parton distributions:
the Sivers function differs by an overall sign in DIS and DY \cite{Col02}.

When integrated over the transverse momentum, one finds the following
distribution from DIS,
\begin{equation}
q_{\rm DIS}(x)
=
\frac{1}{2}
\int \frac{d \xi_-}{2 \pi}
{\rm e}^{- i x \xi_-}
\langle P |
\bar\psi (\xi_-, \bit{0})
[\infty, \bit{0}; \xi_-, \bit{0}]
\gamma^+
[\infty, \bit{0}; 0, \bit{0}]
\psi (0, \bit{0})
| P \rangle
\, ,
\end{equation}
and the distribution from Drell-Yan,
\begin{equation}
q_{\rm DY}(x)
=
\frac{1}{2}
\int \frac{d \xi_-}{2 \pi}
{\rm e}^{- i x \xi_-}
\langle P |
\bar\psi (\xi_-, \bit{0})
[- \infty, \bit{0}; \xi_-, \bit{0}]
\gamma^+
[- \infty, \bit{0}; 0, \bit{0}]
\psi (0, \bit{0})
| P \rangle
\, ,
\end{equation}
processes. The unitarity of the gauge link implies that
\begin{equation}
[\infty, \bit{0}; \xi_-, \bit{0}]
[\infty, \bit{0}; 0, \bit{0}]
=
[- \infty, \bit{0}; \xi_-, \bit{0}]
[- \infty, \bit{0}; 0, \bit{0}]
=
[\xi_-, \bit{0}; 0, \bit{0}]
\, .
\end{equation}
Therefore, both distributions are the same
\cite{Bod85,ColSopSte85,ColSopSte88,ColSopSte89}. In particular, this must
be true in any model calculations.  Therefore, although the approximation
used in the derivation the DY scattering amplitude in Ref.\ \cite{Pei02}
might be invalid for small $\bit{k} \sim x M$, the integrated DY
distribution in a full calculation must reproduce the distribution derived
from DIS.

\section{Conclusion}

In the present study we have found that the transverse momentum-dependent
parton distributions, as defined previously in the literature, are
deficient in light-cone gauges which introduce additional singularities
in perturbative calculations. We have derived their modification which
consists of an extra transverse gauge link at the light-cone infinity.
The regularization of spurious light-cone singularities can be related in
certain cases to boundary conditions on the gauge field, some of which are
consistent with the path integral quantization of the underlying gauge
theory. These boundary conditions shift non-causal interactions, associated
with nature of the gauge potential in the light-like gauge, into the
initial or final state, or their mixture. One can get rid of the final
state interactions in the light-cone gauge $A_+ = 0$ using the advanced
boundary condition for the transverse components of the gauge field. With
this choice, the extra gauge link in DIS parton distributions vanishes.
Therefore, parton distributions are defined solely by the nucleon wave
function and acquire the density interpretation of the naive parton model.
Albeit, in this case the wave function is not real and apart from the
structural information on the hadron it receives also the imaginary phase
mimicking the final state interactions. The density interpretation is lost
for other boundary conditions due to survival of the final state interactions.

The application of these non-causal prescriptions might cause a problem in
explicit computations of multi-loop diagrams and one has to experience
caution using them due to presence of non-integrable pinched singularities.
The latter cancel however in gauge invariant quantities.

Naive universality of transverse momentum-dependent parton distribution in
deeply inelastic scattering and Drell-Yan lepton pair production is generally
violated due to presence of residual gauge factors even in the light-cone
gauge. Setting an advanced or retarded boundary condition, the transverse
link in one of them vanishes, however, it persists in the other. However,
the integrated parton distribution has to obey the universality condition
due to unitary cancellation of gauge effects beyond the interquark separation
points in cross sections.

\vspace{1cm}

We would like to thank I.I. Balitsky and G.P. Korchemsky for enlightening
discussions on several aspects covered in the paper and J.C. Collins,
Yu.V. Kovchegov, A.H. Mueller, S. Peign\'e, G. Sterman and H. Weigert for
useful conversations and correspondence. This work was supported by the US
Department of Energy under contract DE-FG02-93ER40762.

\appendix

\setcounter{section}{0}
\setcounter{equation}{0}
\renewcommand{\theequation}{\Alph{section}.\arabic{equation}}

\section{Maximally fixed light cone gauge}
\label{Residual}

Quantization of a gauge theory is hampered by an infinite redundancy
of gauge degrees of freedom. To solve the problem one has to break
the gauge invariance by imposing a condition on the gauge potential,
$F[A^a_\mu] = 0$. Obvious requirements has to be fulfilled by the
gauge fixing: (i) Every time one sets a constraint, one has to verify
that there exists a choice of the gauge transformation matrix $U$,
\begin{equation}
A_\mu (x) \to A'_\mu (x) = U(x) A_\mu (x) U^\dagger (x)
-
\frac{i}{g} U (x) \partial_\mu U^\dagger (x) \, ,
\end{equation}
that it is realizable in practice. (ii) Next, one cannot gauge away a gauge
covariant quantity like $F^a_{\mu\nu} = \partial_\mu A^a_\nu - \partial_\nu
A^a_\mu - g f^{abc} A^b_\mu A^c_\nu$, so that $F [A^a_\mu] = 0$ must be
consistent with this property.

\begin{figure}[t]
\begin{center}
\hspace{0cm} \mbox{
\begin{picture}(0,145)(100,0)
\put(10,0){\insertfig{6}{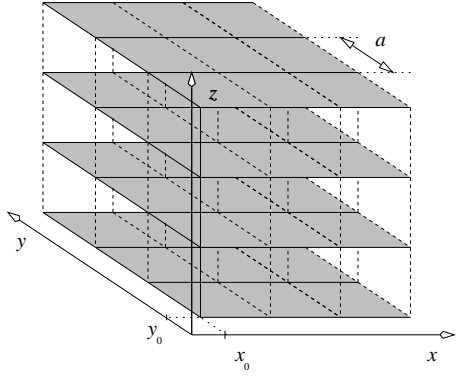}}
\end{picture}
}
\end{center}
\caption{\label{LatticeGaugeFixing} Maximal gauge fixing in three-dimensional
lattice gauge theory. An example of the maximal tree \protect\cite{Cre77,Dun88}:
$A_x (x, y, z) = 0$, $A_y (x_0, y, z) = 0$, $A_z (x_0, y_0, z) = 0$. Full links
are set to $1$ while dashed ones stand for conventional gauge links with non-zero
gluon fields $U_{i,i + a} = \exp\left(- i g a A_i \right)$.}
\end{figure}

Since in the light-cone gauge one ascribes a specific (zero) value to the
light-cone component of the gauge potential, it is very instructive to
consider an equivalent example of temporal gauge fixing in a lattice gauge
theory\footnote{We would like to thank I.I. Balitsky for bringing it to our
attention.} \cite{KogSus75}. Actually, one can consider merely a three-dimensional
gauge theory on a cubic lattice, see Fig.\ \ref{LatticeGaugeFixing}. By a gauge
transformation, one can eliminate one of the components of the gauge
potential, say,
\begin{equation}
\label{Major}
A_x (x, y, z) = 0
\, .
\end{equation}
This sets all link variables along the $x$-direction to unity, solid lines.
This constraint does not fix the gauge completely since we are still free to
perform $x$-independent transformations $U (x_0, y, z)$. To fix the residual
freedom we can fix a subset of links oriented in the two-dimensional
$(y, z)$ plane. By the transformation $U (x_0, y, z)$ we can fix
\begin{equation}
\label{Suppl1}
A_y (x_0, y, z) = 0
\, ,
\end{equation}
but one cannot set at the same time $A_z (x_0, y, z) = 0$ since otherwise
we would set all link variables in the plane $x = x_0$ to unity and this
would imply that we gauged away $F_{yz} (x_0) = 0$. What we can do, however,
is to fix with an $(x, y)$-independent matrix $U (x_0, y_0, z)$ the component
$A_z$ on a line
\begin{equation}
\label{Suppl2}
A_z (x_0, y_0, z) = 0
\, .
\end{equation}
The supplementary conditions (\ref{Suppl1}) and (\ref{Suppl2}) together
with (\ref{Major}) provide the maximal gauge fixing for the gauge theory on
a cubic lattice. Therefore, the maximal gauge fixing is determined by the
maximal tree \cite{Cre77,Dun88}, --- a set of links such that adding one more
results into a closed loop, forbidden by the gauge invariance of the plaquette.

In the continuum gauge theory the above residual conditions are translated
into the boundary conditions on the gauge potential. Obviously, this is not
the only choice for maximal gauge fixing. For instance, in the temporal gauge
the auxiliary Coulomb fixing of residual gauge degrees of freedom
$\vec\partial \cdot \vec A (t_0, \vec x) = 0$ has been explored in Ref.\
\cite{LerMicRos87}. It would be instructive to study a similar constraint for
light-cone gauge $A_+ (x) = 0$ where one may ``freeze" $A_- (L_-, x_+, \bit{x}) = 0$
and $\bit{\nabla} \cdot \bit{A} (L_-, x_+, \bit{x})
= 0$.

It is straightforward to extend the discussion of the lattice example to
the light-cone gauge fixing in QCD. The maximal gauge fixing is achieved
with a set of, e.g., retarded boundary conditions at $L_{-, 1} = - \infty$:
\begin{equation}
\label{RetardBoundary}
A_+ (x_-, x_+, \mbox{\boldmath$x$}) = 0 \, , \quad
A_- (L_-, x_+, \mbox{\boldmath$x$}) = 0 \, , \quad
\mbox{\boldmath$A$}_1 (L_-, 0, \mbox{\boldmath$x$}) = 0 \, , \quad
\mbox{\boldmath$A$}_2 (L_-, 0, L_1, x_2 ) = 0 \, .
\end{equation}
The reason to fix the boundary condition at $x_+ = 0$ is that in the definition
of the parton distributions all fields enter at $x_+ = 0$. The maximally fixed
light-cone gauge is obtained from other gauges by a series of gauge transformations.
First,
\begin{equation}
A_+ (x_-, x_+, \bit{x}) = 0
\end{equation}
with
\begin{equation}
U_+^\dagger (x) = [x_-, x_+, \bit{x}; L_-, x_+, \bit{x}]
\equiv
P \, \exp
\left\{
- i g \int^{x_-}_{L_-} d x'_- A_+ (x'_-, x_+, \bit{x})
\right\} \, .
\end{equation}
At light-cone boundary $L_-$ one performs a gauge transformation on
$A_-$ to nullify it,
\begin{equation}
A_- (L_-, x_+, \bit{x}) = 0
\end{equation}
with
\begin{equation}
U_-^\dagger (x) = [L_-, x_+, \bit{x}; L_-, 0, \bit{x}]
\equiv
P \, \exp
\left\{
- i g \int^{x_+}_{0} d x'_+ A_- (L_-, x_+, \bit{x})
\right\} \, ,
\end{equation}
etc.

Exploiting Eqs.\ (\ref{RetardBoundary}), one can deduce the ``inversion"
formulas to express the gauge potential in terms of the field strength
tensor, namely,
\begin{eqnarray}
A_- (x_-, x_+, \mbox{\boldmath$x$})
\!\!\!&=&\!\!\!
\int_{L_-}^{x_-} d x'_-
[L_-, x_+, \mbox{\boldmath$x$}; x'_-, x_+, \mbox{\boldmath$x$}]
G_{+-} (x'_-, x_+, \mbox{\boldmath$x$})
[x'_-, x_+, \mbox{\boldmath$x$}, L_-, x_+, \mbox{\boldmath$x$}]
\, , \\
\mbox{\boldmath$A$}_1 (x_-, x_+, \mbox{\boldmath$x$})
\!\!\!&=&\!\!\!
\int_{L_-}^{x_-} d x'_-
[L_-, x_+, \mbox{\boldmath$x$}; x'_-, x_+, \mbox{\boldmath$x$}]
G_{+1} (x'_-, x_+, \mbox{\boldmath$x$})
[ x'_-, x_+, \mbox{\boldmath$x$}; L_-, x_+, \mbox{\boldmath$x$}]
\nonumber\\
&+&\!\!\!
\int_{0}^{x_+} d x'_+
[L_-, 0, \mbox{\boldmath$x$}; L_-, x'_+, \mbox{\boldmath$x$}]
G_{-1} (L_-, x'_+, \mbox{\boldmath$x$})
[L_-, x'_+, \mbox{\boldmath$x$}; L_-, 0, \mbox{\boldmath$x$}]
\, , \\
\mbox{\boldmath$A$}_2 (x_-, x_+, \mbox{\boldmath$x$})
\!\!\!&=&\!\!\!
\int_{L_-}^{x_-} d x'_-
[L_-, x_+, \mbox{\boldmath$x$}; x'_-, x_+, \mbox{\boldmath$x$}]
G_{+2} (x'_-, x_+, \mbox{\boldmath$x$})
[ x'_-, x_+, \mbox{\boldmath$x$}; L_-, x_+, \mbox{\boldmath$x$}]
\nonumber\\
&+&\!\!\!
\int_{0}^{x_+} d x'_+
[L_-, 0, \mbox{\boldmath$x$}; L_-, x'_+, \mbox{\boldmath$x$}]
G_{-2} (L_-, x'_+, \mbox{\boldmath$x$})
[L_-, x'_+, \mbox{\boldmath$x$}; L_-, 0, \mbox{\boldmath$x$}]
\nonumber\\
&+&\!\!\!
\int_{L_1}^{x_1} d x'_1
[L_-, 0, L_1, x_2; L_-, 0, x'_1, x_2]
\nonumber\\
&&\qquad\qquad\qquad\quad\ \times G_{12} (L_-, 0, x'_1, x_2)
[L_-, 0, x'_1, x_2; L_-, 0, L_1, x_2]
\, .
\end{eqnarray}
For $x_+ = 0$, the transverse components simplify,
\begin{eqnarray}
\mbox{\boldmath$A$}_1 (x_-, 0, \mbox{\boldmath$x$})
\!\!\!&=&\!\!\!
\int_{L_-}^{x_-} d x'_- G_{+1} (x'_-, 0, \mbox{\boldmath$x$})
\, , \nonumber\\
\mbox{\boldmath$A$}_2 (x_-, 0, \mbox{\boldmath$x$})
\!\!\!&=&\!\!\!
\int_{L_-}^{x_-} d x'_- G_{+2} (x'_-, 0, \mbox{\boldmath$x$})
+
\int_{L_1}^{x_1} d x'_1 G_{12} (L_-, 0, x'_1, x_2)
\, ,
\end{eqnarray}
and we dropped the path-ordered exponentials. Using these results we
can compute the gauge propagator. For instance,
\begin{eqnarray}
\label{ComponentsMinusTrans}
\langle 0 | T \left\{
A_- (x_-, x_+, \mbox{\boldmath$x$})
\mbox{\boldmath$A$}_i (y_-, 0, \mbox{\boldmath$y$})
\right\} | 0 \rangle
\!\!\!&=&\!\!\!
\int_{L_-}^{x_-} d x'_-
\int_{L_-}^{y_-} d y'_-
\langle 0 | T \left\{
G_{+-} (x'_-, x_+, \mbox{\boldmath$x$})
G_{+ i} (y'_-, 0, \mbox{\boldmath$y$})
\right\} | 0 \rangle
\nonumber\\
&=&\!\!\!
i \int \frac{d^4 k}{(2 \pi)^4} {\rm e}^{- i k \cdot (y - x)}
\frac{1}{k^2 + i 0} \frac{\bit{k}_i}{k_+ + i 0} \, ,
\end{eqnarray}
where we have used the fact that the second term $G_{12}$ in $\bit{A}_2$ does
not work. Analogously, we get for other components
\begin{eqnarray}
\label{ComponentsTransMinus}
\langle 0 | T \left\{
\bit{A}_i (x_-, 0, \bit{x})
A_- (y_-, y_+, \bit{y})
\right\} | 0 \rangle
\!\!\!&=&\!\!\!
i \int \frac{d^4 k}{(2 \pi)^4} {\rm e}^{- i k \cdot (y - x)}
\frac{1}{k^2 + i 0} \frac{\bit{k}_i}{k_+ - i 0} \, ,
\nonumber\\
\langle 0 | T \left\{
A_- (x_-, x_+, \bit{x})
A_- (y_-, y_+, \bit{y})
\right\} | 0 \rangle
\!\!\!&=&\!\!\!
i \int \frac{d^4 k}{(2 \pi)^4} {\rm e}^{- i k \cdot (y - x)}
\frac{2 k_-}{k^2 + i 0} {\rm PV} \frac{1}{k_+} \, ,
\end{eqnarray}
where PV stands for the principal value prescription
\begin{equation}
{\rm PV} \frac{1}{k_+}
\equiv
\frac{1}{2}
\left\{
\frac{1}{k_+ + i 0}
+
\frac{1}{k_+ - i 0}
\right\} \, .
\end{equation}
This is actually the only components which contribute in practical
calculation of section \ref{LightConeGaugeComputations}.

\begin{figure}[t]
\begin{center}
\hspace{0cm} \mbox{
\begin{picture}(0,30)(45,0)
\put(10,0){\insertfig{2}{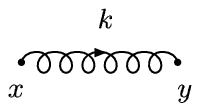}}
\end{picture}
}
\end{center}
\caption{\label{GluonPropagator} Feynman rules for the gluon propagator.}
\end{figure}

If we delinquently extend the results (\ref{ComponentsMinusTrans}) and
(\ref{ComponentsTransMinus}) out of the plane $x_+, y_+ = 0$, which implies
that we exceed the amount of gauge fixing freedom in setting the boundary
condition, we get the light-cone propagator
\begin{equation}
\label{AdvancedGluonPropagator}
\langle 0 | T \left\{
A_\mu (x)
A_\nu (y)
\right\} | 0 \rangle
=
- i \int \frac{d^4 k}{(2 \pi)^4} {\rm e}^{- i k \cdot (y - x)}
\frac{1}{k^2 + i 0}
\left(
g_{\mu \nu}
-
\frac{k_\mu n_\nu}{k_+ - i 0}
-
\frac{k_\nu n_\mu}{k_+ + i 0}
\right)
\, .
\end{equation}
The gauge overfixing does not lead to complications in perturbation theory,
since it boils down to the assumption $G_{\mu\nu} (x_- \to - \infty, x_+, \bit{x})
\to 0$. The direction of momentum flow $k$ in Eq.\ (\ref{AdvancedGluonPropagator})
is assumed to be from the point $x$ to $y$, see Fig.\ \ref{GluonPropagator}. The
very same propagator was derived previously in the framework of path integral by
means of a gauge transformation from the temporal gauge $A_0 = 0$ to the regularized
gauge $A_+ + \varepsilon \partial_+ A_+ = 0$ ($\varepsilon \to \pm 0$) \cite{SlaFro88}
and canonical quantization in Ref.\ \cite{Ant88}. It was used more recently in the
literature in the context of specific computations, see, e.g., \cite{Kov97,FerIanLeoMcl01}.

If one uses the light cone gauge advanced boundary conditions
$L_{-, 1} = \infty$:
\begin{equation}
\label{AdvancedBoundary}
A_+ (x_-, x_+, \mbox{\boldmath$x$}) = 0 \, , \quad
A_- (L_-, x_+, \mbox{\boldmath$x$}) = 0 \, , \quad
\mbox{\boldmath$A$}_1 (L_-, 0, \mbox{\boldmath$x$}) = 0 \, , \quad
\mbox{\boldmath$A$}_2 (L_-, 0, L_1, x_2 ) = 0 \, ,
\end{equation}
the contour deformation is moved into the opposite half-plane as compared
to Eqs.\ (\ref{ComponentsMinusTrans}) and (\ref{ComponentsTransMinus}).
With the same reasoning as preceded Eq.\ (\ref{AdvancedGluonPropagator}),
one gets
\begin{equation}
\langle 0 | T \left\{
A_\mu (x)
A_\nu (y)
\right\} | 0 \rangle
=
- i \int \frac{d^4 k}{(2 \pi)^4} {\rm e}^{- i k \cdot (y - x)}
\frac{1}{k^2 + i 0}
\left(
g_{\mu \nu}
-
\frac{k_\mu n_\nu}{k_+ + i 0}
-
\frac{k_\nu n_\mu}{k_+ - i 0}
\right)
\, .
\end{equation}

The conventional antisymmetric combination of boundary conditions
(\ref{RetardBoundary}) and (\ref{AdvancedBoundary}) boundary conditions,
schematically,
\begin{equation}
\bit{A} (\infty) + \bit{A} (- \infty) = 0
\end{equation}
are not consistently implementable within the path integral approach.
Moreover, one cannot find the corresponding gauge transformation which
would result into this residual gauge fixing. However, if one accepts
it in the ``inversion'' formulas, it results into the the propagator
with $1/k_+$ singularity regularized via the principal value prescription.


\end{document}